%% file: main.tex
\documentclass[11pt,a4paper,table]{article}

\usepackage[utf8]{inputenc}
\usepackage{graphicx}
\usepackage{comment}
\usepackage{multirow}
\usepackage{enumitem}
\usepackage{setspace}
\usepackage{amsfonts}
\usepackage{amsmath,amsthm,amssymb}
\usepackage{float}
\usepackage{tikz}
\usepackage{subcaption}
\usepackage{natbib}
\usepackage{lscape}
\usepackage{rotating}
\usepackage[normalem]{ulem}
\usepackage{authblk}

\usepackage[margin=1in]{geometry}

\newtheorem{thm}{Theorem}

\newtheorem{hyp}{Hypothesis}
\newtheorem{rem}{Remark}

\title{Rankings-Dependent Preferences: A Real Goods Matching Experiment\footnote{Emails: akloosterman@iso-ne.com and troyan@virginia.edu. For their comments, we would like to thank Vincent Meissner, Assaf Romm, Yingzhi Liang, Alex Rees-Jones, and audiences at the 12th Conference on Economic Design in Padova and the 2023 WZB Berlin Matching Market Design Conference. We gratefully acknowledge support from the UVA Buckner W. Clay Dean of Arts and Sciences, Vice President for Research, and Quantitative Collaborative. This project was approved by the Institutional Review Board of the University of Virginia (Protocol Number 5421).}}
\author[1]{Andrew Kloosterman}
\author[2]{Peter Troyan}
\affil[1]{ISO New England}
\affil[2]{Department of Economics, University of Virginia}

\date{August 2024}

\begin{document}

\maketitle

\begin{abstract}
We investigate whether preferences for objects received via a matching mechanism are influenced by how highly agents rank them in their reported rank order list.  We hypothesize that all else equal, agents receive greater utility for the same object when they rank it higher.   The addition of rankings-dependent utility implies that it may not be a dominant strategy to submit truthful preferences to a strategyproof mechanism, and that non-strategyproof mechanisms that give more agents objects they \emph{report} as higher ranked may increase market welfare. We test these hypotheses with a matching experiment in a strategyproof mechanism, the random serial dictatorship, and a non-strategyproof mechanism, the Boston mechanism.  A novel feature of our experimental design is that the objects allocated in the matching markets are real goods, which allows us to directly measure rankings-dependence by eliciting values for goods both inside and outside of the mechanism. The experimental results are mixed, with stronger evidence for rankings-dependence in the RSD treatment than the Boston treatment.   We find no differences between the two mechanisms for the rates of truth-telling and the final welfare.
\end{abstract}
\pagebreak

\section{Introduction}

In strategyproof mechanisms, it is always an optimal strategy for agents to truthfully report their private information to the mechanism.  This theoretical property is clearly appealing, as it gives a mechanism designer the ability to predict play and make meaningful statements about other criteria such as welfare.  However, a growing body of empirical evidence has documented significant deviations from truthful reporting in such mechanisms.  This issue is particularly important in matching markets, the focus in this paper, in which participants submit preference rankings of alternatives such as schools or medical residency programs to a centralized clearinghouse which determines the assignment. Evidence of non-truthful behavior in strategyproof matching mechanisms can be found both in the lab \citep{chen/sonmez:06,pais_pinter08,li2017obviously} and in high-stakes decisions in the field \citep{chen2019self,shorrer2018obvious,hassidim2021limits}.

Deviations from truthful reporting are harmful under the implicit assumption that agents' preferences are standard economic preferences in that the values for the objects they receive in the mechanism are determined solely by the characteristics of these objects.  If this assumption holds, then, when we observe agents rank objects they value less above objects they value more in a strategyproof mechanism, we can claim that these deviations from truthful reporting are indeed ``mistakes".  But what if this underlying assumption is wrong?  Then, these mistakes may not really be mistakes, but rather optimal behavior from agents with non-standard preferences.

In this paper, we explore the possibility that the rankings agents submit to the mechanism influence values.  For instance, an agent may value an object higher when they rank it $2^{nd}$ compared to a counterfactual in which they receive the same object but rank it $4^{th}$, because they suffer disutility when they get a low-ranked object.  There are a number of reasons why receiving low-ranked objects may be undesirable. These include reference-dependent loss aversion where agents expect to get a high-ranked object and are disappointed when they do not \citep{dreyfuss2019expectations,meisner2023report}, ego utility where agents think that it looks good to others to receive a high-ranked object \citep{koszegi2006ego}, preferences that focus on beating others rather than maximizing one's own utility \citep[a ``joy of winning",][]{cooper2008understanding}, or limited information on quality that instigates a `curse of acceptance' whereby receiving a low-ranked object indicates that it is bad \citep{kloosterman2020school}.  

Following this discussion, we consider agents who have utility from receiving object $x$ that takes the form
\[
u(x)= v(x)+\rho(\text{rank}(x))
\]
We call $v(x)$ the agent's \emph{fundamental value} for object $x$; this corresponds to the standard economic preferences assumed in typical matching models. The second term, $\rho(\text{rank}(x))$, is an additional \emph{rankings-dependent utility} component that is determined by how highly the agent ranked $x$ in their reported preferences. When $\rho(j)=0$ for all $j$, the model reduces to a standard model of preferences. In this paper, we investigate the possibility that $\rho$ may not be identically zero, and in particular, that it is a decreasing function; in other words, all else equal, agents receive more utility when they rank an object higher.

Investigating questions of rankings-dependent utility is important because if rankings-dependent utility is an extant phenomenon, there are consequences for real-world market design. Indeed, the main motivation for this paper was discussions with school district administrators who continue to prefer the Boston mechanism over strategyproof alternatives such as deferred acceptance, because Boston gives more students their reported first choices. The standard critique is that this data cannot be taken at face-value, because the Boston mechanism gives clear incentives for agents to manipulate their preferences, even in the absence of rankings dependence (see \cite{dur2018identifying} for evidence of such behavior in a real-world school choice environment).  The response from administrators is that while they understand this argument in theory, it is missing an important issue in practice:  \emph{parents just do not like to get something they ranked low in their list} (Cambridge, MA School District, personal communication). If this is correct, then this should be incorporated into matching market models to better reflect  how real-world agents make decisions in mechanisms and evaluate outcomes.

Thus, the main contribution of this paper is to test this hypothesis: we design and implement a experiment to explore the existence (or lack thereof) of rankings-dependent utility. This would be nearly impossible to do in a field setting, as people usually participate in a mechanism only once and it would be impossible to disentangle their fundamental value from any potential rankings-dependent utility. Thus, we use a laboratory experiment, which will allow us to do precisely that.

The mechanisms that we use for our experiment are the random serial dictatorship (RSD) and the Boston mechanism.  We chose these mechanisms because they are two canonical mechanisms that are widely used in practice.\footnote{Deferred acceptance \citep[DA, ][]{gale/shapley:62} is another popular mechanism used in practice. RSD is a special case of DA where the priorities at each object are equivalent (and random). We chose RSD and Boston because these are simpler mechanisms to explain than DA.} Further, RSD is strategyproof, while the Boston mechanism is not, yet the Boston mechanism may result in agents receiving higher-ranked goods. This allows us to answer not only our main question of rankings-dependent utility, but also to test the hypothesis that a non-strategyproof mechanism may be welfare-enhancing once rankings-dependent utility is taken into account. 

A novel and key feature of our experimental design is that we use real objects that are in the room at the time of the experiment and that the participants may take home with them. To determine whether utility is rankings-dependent, in Phase I of the experiment, we first elicit valuations for 20 common objects (backpacks, alarm clocks, phone chargers, etc.) with the multiple price list (MPL) elicitation method. In Phase II of the experiment, five of the objects (a Fjallraven backpack, a Hydroflask water bottle, a Moleskine notebook, a generic ceramic coffee mug, and a package of 4 ballpoint pens) were chosen and the participants were asked to submit a rank-order list of these five objects to a mechanism. The mechanism (either RSD or Boston, depending on the treatment) produces an allocation of one object to each participant. After the mechanism, we once again elicit each participant's valuation for the object that they were allocated in the mechanism. The Phase I valuation measures the fundamental value $v(x)$ independent of any mechanism, while the Phase II valuation measures $v(x)+\rho(j)$, the value for $x$ when it is received in a mechanism in which it is ranked $j^{th}$. So, the difference---which we call Net Value (NV)---is a measure of $\rho$, the rankings-dependent utility term:
$$
NV(j) = \text{Phase II value }-\text{ Phase I value}=(v(x)+\rho(j))-v(x)=\rho(j).
$$
 Under the hypothesis that utility is not rankings-dependent ($\rho=0$), we should see no difference in valuations between the two phases; under the alternative hypothesis, $\rho$, and hence the difference in valuations, is a non-zero function that is decreasing in the rank of the object received.\footnote{It is important to emphasize that while MPL elicitation may itself be susceptible to certain behavioral ``biases'' (see, e.g., \cite{andersen2006elicitation}), we use the \emph{same} elicitation method in both phases and compare \emph{differences} in valuations across phases to compute Net Value. Thus, nonzero Net Values cannot be attributed to the choice of elicitation method. We elaborate on this point in Section \ref{sec:Experimental-Design}.}

The use of real goods is a significant departure from the experimental literature on matching mechanisms which usually uses fictitious ``goods'' with induced monetary values (\cite{hakimov2021experiments} provide a recent survey of this literature). An induced values design is insufficient for our purposes because rankings-dependent preferences, by construction, are subjective so we need to elicit them.  We use a standard elicitation mechanism, the multiple price list, that elicits this subjective utility in dollar terms.  Monetary prizes would not work because we would be eliciting the value of money in dollar terms and it is almost certain that we would find that people value money at its monetary value. We would therefore conclude that $\rho(\text{rank}(x))$ would be zero because of this improper elicitation. The use of real goods also more faithfully simulates real-world environments, where participants must form their own preferences rather than have them being induced.\footnote{The only other example of real goods in a matching experiment we are aware of is \cite{guillen2018effectiveness}, who conduct a field experiment to study the impact of top-down advice on truth-telling rates in a classroom setting in which students were assigned to one of three term paper topics. Similar to us, they faced the issue of how to elicit subjects' true preferences outside of the mechanism, so they could compare to behavior inside of the mechanism. They solved this problem by first having students believe they could choose any of the three topics and asking them to indicate their favorite. Later in the semester, they ``surprised'' the students   and told them that 1/3 of the class had to be assigned to each of the three topics, and the students were asked to submit a rank list. The final allocation was determined using the top trading cycles mechanism. Notice that this method elicited an ordinal preference for the top choice only.} 

Our main hypothesis of interest is that the NV---which measures $\rho(\text{rank}(x))$---should be non-zero, and in particular decreasing in the reported rank of the good received in the mechanism. For the RSD treatment, NV is nearly monotonically decreasing, from an average of +\$2.87 for participants who receive their top-ranked good to an average of $-\$0.69$ for participants who receive their fifth-ranked good (out of five). For the Boston treatment, on the other hand, there is a much smaller increase in NV for the first ranked good (only about +\$0.60), and, looking at the raw data averages, there is no clear evidence that NV is decreasing in rank.  Non-parametric tests provide statistical support for these impressions. Moving to regression analysis, we find some support for the hypothesis in both treatments, with the rank being a statistically significant predictor of NV for both the RSD treatment and the Boston treatment separately, as well as for the pooled sample. Interestingly, both the Phase I value and a risk aversion regressor (measured using the Holt-Laury switching point) are strongly statistically significant in the Boston treatment, but not in the RSD treatment.

Beyond simply exploring the possible existence of rankings-dependent preferences, we also study some of the implications they may have in terms of truth-telling and welfare. With the introduction of rankings dependence into preferences, truth-telling is no longer an equilibrium of either RSD or Boston. This is because participants now have an incentive to rank less popular goods higher under both mecahnisms, to achieve a higher rankings-dependent utility, and solving for an equilibrium becomes very complex. In Section 3, we provide some theoretical results under certain simplifying assumptions that are relevant for the experiment.  In particular, we show that in equilibrium, relative to RSD, the Boston mechanism has (i) weakly less truthful reporting (Theorem \ref{thmtt}), (ii) weakly more agents getting their top-ranked object (Theorem \ref{thm-first-gooods-first}).  and (iii) weakly higher welfare in Boston (Theorem \ref{thmwel}).

In the experiment, we find no differences in the rates of truth-telling between the two treatments according to a wide range of measures. This is interesting given the plethora of experimental work going back to \cite{chen/sonmez:06} that generally finds much less truth-telling in Boston than in deferred acceptance (which is equivalent to RSD in our setting). In our experiment, this is driven by much lower rates of truth-telling in RSD relative to the literature (as opposed to higher rates of truth-telling in Boston).\footnote{These comparisons are not perfect, as most of this literature studies DA, while we study RSD, which is a special case of DA. A survey of matching experiments by \cite{hakimov2021experiments} reports truth-telling rates in the (arguably more complex) DA mechanism in the range of 55-85\%, though this varies with the details of the treatment. \cite{li2017obviously} does study RSD specifically, and finds truth-telling rates of about 60\%,  compared with 40\% for our experiment.  All of these experiments use an induced-values design.}   Note that this is what would be expected if preferences are indeed rankings-dependent: reporting truthfully (i.e., according to elicited Phase I values) is no longer an optimal strategy, and so the truth-telling rates in RSD should be lower than in other experiments.\footnote{While the lower rates of truth-telling in RSD is an interesting result, and is consistent with our theory, the lack of differences across treatments is less significant, and could just be an artifcat of the details of our experiment, and in particular the goods that were chosen. (Indeed, our theoretical results only say that RSD should have \emph{weakly} higher rates of truth-telling, which is consistent with the data.) Further, the use of real goods also may complicate comparisons with previous studies: with induced values, if a participant does not understand the mechanism, a natural default may be to just rank the fictitious goods according to the values given, which would lead to higher rates of truth-telling in induced values environments. Thus, while induced values provide greater levels of control for the experimenter, the results from such experimental designs may not carry over to real-world settings where actual goods are being distributed.}

To summarize, we find mixed evidence that utility is rankings-dependent in matching mechanisms. Data from the RSD treatmeant is consistent with the hypothesis, while evidence from the Boston treatmeant is less clear-cut, though we do find rank to be a significant predictor after controlling for subjects' risk aversion.  We find no differences in welfare in our experiment, but this will depend on the details of the preference environment: we are the first (to our knowledge) to show theoretically that in some settings, rankings-dependent utility can upend the standard welfare comparisons between mechanisms. We also find much lower rates of truth-telling in the strategyproof RSD mechanism compared to the literature, which is consistent with rankings-dependent preferences, as the RSD mechanism is no longer strategyproof.\footnote{In general, solving for the equilibrium is very difficult once rankings-dependence preferences are introduced, and our theoretical results hold only under particular assumptions on the preference environment. While we attempted to replicate this in the experiment as much as possible, the use of real goods instead of induced values inherently gives us less control over the preferences. Thus, the lack of differences given the specific objects and design of our experiment does not preclude differences in other environments.}

A further contribution of our paper is the use of real goods in matching market experiments. This design choice was necessitated by the nature of rankings-dependent preferences we are trying to study, but we think it may be useful for other researchers in the experimental matching literature, which as to now has almost exclusively used induced values designs. While the use of real goods certainly complicates some aspects of experimental design and analysis, it also more faithfully simulates real-word markets, where agents must form their preferences rather than having them given to them. To the extent that preference formation plays a role in how agents participate in matching mechanisms---whether via rankings-dependence as we model it, or other channels---real goods experiments will be important in understanding the outcomes of matching mechanisms. We provide some initial evidence of such a phenomenon in a simple setting (RSD), though much more work is needed to more thoroughly understand how agents form preferences in matching settings. 

\begin{subsubsection}*{Related Literature}
 One of the main motivations for this paper is the growing body of evidence of ``mistakes'' in strategyproof mechanisms (see the first paragraph of the introduction for references). There are several ways one can proceed from these observations. Assuming that the underlying preference model is correct, and these mistakes are actually mistakes, one response is that the designer should simply invest more in communicating how the mechanism works and teaching the participants that truthful reporting is in their interest. \cite{rees2018experimental} conduct a ``lab-in-the-field'' experiment in which they recruit medical students who have just gone through the National Resident Matching Program (NRMP) to participate in a related lab experiment, and find a significant fraction of participants did not report truthfully, despite having just participated in the same mechanism in a high stakes environment in which the NRMP invests heavily in tutorials describing how the mechanism works. If this approach is unsuccessful, another possibility is to design mechanisms that are more easily recognizable as strategyproof by the participants on their own. For instance, \cite{li2017obviously} introduces the notion of obvious strategyproofness as a desideratum for mechanism design, with the idea being that if the mechanism is designed to satisfy this criterion, the players will be able to recognize themselves that truthful reporting is a dominant strategy, and will be more likely to report truthfully. Li's paper has led to a rapidly expanding literature on obvious mechanism design as a way to limit mistakes by participants.\footnote{Theoretical explorations include \cite{ashlagi-gonczarowski-OSP}, \cite{troyan2016obviously}, \cite{pycia-troyan-simplicity}, and \cite{bade-gonczarowski-2016}. For lab experiments, see \cite{zhang2017partition} and \cite{bo-hakimov-2019}.}

We take a different approach in this paper, which is a reassessment of the assumption that agent preferences are determined solely by the characteristics of the goods they receive. We discuss a few other recent papers that have explored related ideas. 

The closest paper to ours theoretically is \cite{meisner2023report}. He proposes an equivalent model of utility that consists of both a fundamental value plus a rankings-dependent component. He then focuses on strategyproof mechanisms, and proves that any non-truthful preference ranking can be rationalized as optimal for some beliefs over match probabilities (what he refers to as ``attainability distributions'', which are determined by the mechanism itself combined with beliefs about the strategies of the other agents).

\cite{meisner2021school}, \cite{dreyfuss2019expectations}, \cite{dreyfuss2022deferred}, and \cite{chen2023people}  all focus specifically on expectations-based reference-dependent preferences (EBRD preferences for short, also referred to as EBLA for expectations-based loss aversion; \cite{kHoszegi2006model}) as a possible explanation for seemingly dominated choices in strategyproof mechanisms. \cite{meisner2021school} introduce loss aversion into school choice problems and theoretically study its implications for submitting non-truthful rank order lists. \cite{dreyfuss2019expectations} re-evaluate the experimental data of \cite{li2017obviously}, who finds mistakes in the non-obviously strategyproof RSD mechanism, and find that EBRD preferences might explain this behavior. 

\cite{dreyfuss2022deferred}  build a matching model with EBRD preferences and conduct a lab experiment using four different implementations of the DA mechanism: static (student) proposing, static receiving, dynamic proposing, and dynamic receiving. Using a model of loss aversion, they derive predictions on the relative proportion of non-truthtelling behavior across these four mechanisms.\footnote{What we call non-truthtelling behavior they refer to as ``non-straightforward behavior''.} They show that, for plausible values of the loss aversion parameter, the trends in the data qualitatively match those predicted by the theory; that is, they find more truthtelling in the treatments that are predicted to have higher rates of truthtelling according to the theory.  

While similar in motivation, the paper by \cite{dreyfuss2022deferred} is very different from ours, from both a methodological and experimental design perspective. They consider a particular utility model (EBRD) and show that the trends across treatments in the data qualitatively match the trends predicted by EBRD. While these trends are suggestive evidence for EBRD preferences, in terms of overall levels, they also find much more non-truthtelling behavior than the EBRD predictions, and write that ``the EBRD model, while explaining a lot of the observed data, appears to be an incomplete explanation''. For instance, another explanation could be that subjects misunderstand the game they are playing. Because of this potential confound, \cite{chen2023people} design a much simpler experimental environment to eliminate the possibility of game form misunderstanding, but which still predicts non-truthtelling behavior for loss-averse subjects. In this simplified environment, they find very high rates of truth-telling and no evidence that loss aversion drives misreporting behavior. This leads \cite{chen2023people} to argue that loss aversion is likely not a relevant explanation for mistakes in matching mechanisms.\footnote{In addition to risk aversion, we also had our subjects complete a standard loss aversion task. We did not find loss aversion to be a significant predictor in any of our regressions.} Note that both \cite{dreyfuss2022deferred} and \cite{chen2023people} use induced values designs, so neither experiment would be able to capture other aspects of rankings-dependent preferences that would only be found with a real-goods design, as in our paper.

Our experiment, on the other hand, was designed to 
provide a direct measurement of non-standard preferences. While we are agnostic on the underlying source, we think this is a feature, rather than a bug, of our design: by taking a direct measurement of non-standard preferences, we avoid the potential confounds that arise in trying to indirectly infer their existence from qualitative trends across treatments that would be predicted by a particular theory, as discussed above. We think this makes our paper a useful complement to the other approaches that have been taken. Further, both \cite{dreyfuss2022deferred} and \cite{chen2023people} use induced values and reduce the game to an individual decision problem where participants are effectively asked to choose between objective lotteries. We use real goods, and have participants play in a multiplayer game. This is important to the extent that factors such as preference formation itself over objects and relative comparisons of one's own outcome against others influence agent preferences, issues which are  likely to be present in real-world settings. While the goal of the present paper was more modest---to simply evaluate the existence of non-standard preferences---we think that a natural direction for future work is deeper analysis of the microfoundations behind them.

 \end{subsubsection}

\section{Model}\label{sec:model}

\subsection{Preferences and Mechanisms}

There is a set $I=\{i_1,i_2,\ldots,i_N\}$ of  agents and a set $X=\{x_1,x_2,\ldots,x_N\}$ of objects. A \textbf{matching} is a function $\mu:I\rightarrow X$ where $\mu(i)\in X$ is the object that is assigned to agent $i$. There are an equal number of agents and objects, and so we assume that all matchings assign a unique object to every agent, i.e., if $i\neq j$, then $\mu(i)\neq \mu(j)$. Let $\mathcal{M}$ denote the set of matchings. Notice that we assume each object has capacity 1, an equal number of agents and objects, and the objects have no preferences or priorities over the agents. All these restrictions can easily be generalized to capture important features of real-world settings such as school choice. We make these assumptions because they are all that is needed to study our main phenomena of interest, and they ensure that the model corresponds as directly as possible to the experiment that we run.

Let $\mathcal{P}$ be the set of all strict ordinal rankings over $X$. For any $P_i\in \mathcal{P}$, we write $xP_i y$ to denote that agent $i$ strictly prefers $x$ to $y$. We use $R_i$ for the corresponding weak relation, i.e., $xR_i y$ if either $xP_i y$ or $x=y$. A \textbf{mechanism} is a function $\psi:\mathcal{P}^N \rightarrow \mathcal{M}$. We write $\psi_i(P)\in X$ for the object allocated to $i$ at preference profile $P = (P_1,\ldots,P_N)$.  Every mechanism induces a game in which the action space for each agent is $\mathcal{P}$.  

In most of the matching literature, it is assumed that an agent's utility is determined solely by the object received. We deviate from this assumption by allowing an agent's utility to depend on both (i) the object received and (ii) the position in which they ranked the object in their preferences. Formally, agent $i$'s utility from submitting a reported preference ranking $P_i$ and receiving object $x$ is 
\begin{equation}\label{eq:rank-dep-utility-function}
u_i(x,P_i) = v_i(x) + \rho(j).     
\end{equation}
where $j=|\{x'\in X: x'R_i x\}|$ is the rank of object $x$ in the reported preference list. We call  $v_i(x)$ agent $i$'s \textbf{fundamental value} for object $x$, and $\rho(j)$ agent $i$'s \textbf{rankings-dependent utility} from receiving the object that she ranked in the $j^{th}$ position.  Note that in this formulation, the function $\rho(\cdot)$ is the same for all agents. This could be generalized to allow for heterogeneity, but given our experimental design, it is infeasible to measure a different $\rho(j)$ for each agent and each $j$, and so we omit this generalization in the model.

The main assumption of our model is that, all else equal, participants will prefer to get objects when they rank them higher in their submitted preferences.  For example, receiving good $x$ provides more utility when it is ranked $2^{nd}$ than when it is ranked $4^{th}$. Hence, we assume the function $\rho$ is such that $\rho(1)\geq\rho(2)\geq\cdots \geq\rho(n)$. We discussed in the introduction possible microfoundations for rankings-dependent utility. In the experiment, we do not attempt to discern between these explanations, but seek simply to determine whether preferences do in fact depend on rankings. 

For our experiment, we consider two mechanisms which we describe here: random serial dictatorship and the Boston mechanism.

\subsubsection*{Random Serial Dictatorship (RSD)}

The random serial dictatorship mechanism works as follows. Each agent submits a strict ordinal ranking over all of the objects. The mechanism then draws an ordering of the agents randomly from the uniform distribution over all possible agent orderings, and proceeds in rounds as follows:
\begin{itemize}
    \item Round 1: The first agent in the order is assigned their top-ranked object according to their submitted preference ranking.
    \item Round 2: The second agent in the order is assigned their top-ranked object that was not assigned in the first round.
    \item Round $k=3,\ldots,N$: The $k^{th}$ agent in the order is assigned their top-ranked object that was not assigned in any earlier round $1,\ldots,k-1$. 
\end{itemize}

\subsubsection*{Boston Mechanism}
The Boston mechanism works as follows. Each agent submits a strict ordinal ranking over all of the objects. The mechanism draws an ordering of the agents randomly from the uniform distribution over all agent orderings, which will be used as a ``tie-breaker'' below.  The mechanism then proceeds in rounds as follows:
\begin{itemize}
    \item Round 1: The mechanism considers the top-ranked object of each agent. If only one agent has ranked an object first, the object is assigned to that agent. If more than one agent has ranked an object first, the mechanism assigns the object to the agent among them who was ranked highest in the random agent ordering drawn above. The agents and objects that were assigned leave the market. If all agents have been assigned an object, the mechanism ends. Otherwise, all agents and objects that were not assigned in this round proceed to the next round. 
    \item Round 2: Only agents who were not assigned an object in the first round participate. The mechanism looks at the second-ranked choices of all such agents. If only one agent has ranked an object second, the object is assigned to that agent. If more than one agent has ranked an object second, the mechanism assigns the object to the agent among them who was ranked highest in the random agent ordering drawn above.  The agents and objects that were assigned leave the market. If all agents have been assigned an object, the mechanism ends. Otherwise, all agents and objects that were not assigned in this round proceed to the next round. 
    \item Round $k=3,\ldots,M$: The mechanism proceeds exactly as in round 2, except it now considers the $k^{th}$-ranked objects in the rank lists of the agents who participate in this round.
\end{itemize}

The definitions of RSD and Boston given above are simplifications of a more general class of mechanisms. It is easy to extend the mechanisms to incorporate features such as multiple copies of each object, outside options, and priority lists over the agents for each object, all of which are common in settings such as school choice (see, for instance, \cite{abdul2003school}). We use these definitions because they are sufficient to capture the properties of assignment mechanisms that are the focus of our investigation while still remaining parsimonious enough to design a lab experiment to cleanly test our hypotheses of rankings-dependent preferences.

\subsection{Incentives and Welfare}

Two key concerns when designing any allocation mechanism are the incentives they provide participants with regard to reporting their preferences and the equilibrium welfare of the resulting allocation. In this section, we briefly discuss these issues for both RSD and Boston.  After discussing the details of the experimental design, we will return to these issues with formal theorems and experimental hypotheses based on them.

\subsubsection*{Truth-telling}

A common desideratum when designing a matching mechanism is strategyproofness, the property that it is a weakly dominant strategy in the mechanism-induced game for each agent to submit their true preference ranking to the mechanism.  In the standard matching literature where behavior is not rankings-dependent (i.e., $p(j)=0$ for all $j$), this corresponds to submitting the preference ranking that lists objects in decreasing order of fundamental value. This strategy is, for obvious reasons, called the \emph{truthful strategy}.  Without rankings-dependent utility,  RSD is well-known to be strategyproof while Boston is not. Intuitively, in Boston an agent with fundamental values $v_i(x)>v_i(y)$ may want to lie and report $yP_i x$, if $x$ is likely to be very popular while $y$ is almost as good and easier to get, because failing to get $x$ might lead to getting an even worse object $z$. Indeed, \cite{troyan2020obvious} show that not only is Boston manipulable, it is \emph{obviously} manipulable, and there is empirical evidence from school choice data that show families engaging in precisely this type of non-truth-telling behavior \citep{dur2018identifying}. 

With the introduction of rankings-dependent utility in Equation (\ref{eq:rank-dep-utility-function}), the strategyproofness property is not well-defined because there is no one-to-one mapping of utility to a preference ranking of the objects.  Nevertheless, it is straightforward to still define the truthful strategy with respect to fundamental values exactly as above as the strategy that submits a preference ranking that lists objects in decreasing order of fundamental value.  Though this terminology is not as descriptive in this model where fundamental value does not reflect utility, we continue to use it to maintain consistency with the previous literature. Due to the rankings-dependent utility, it is no longer the case that choosing the truthful strategy is weakly dominant in RSD.  There is extensive empirical and experimental evidence (see the first paragraph in the introduction for references) that agents do not submit the truthful strategy to the mechanism in deferred acceptance, a mechanism that is equivalent to RSD in our setup. Indeed, this is the main motivation for incorporating and investigating rankings-dependent utility in the RSD mechanism.

\subsubsection*{Efficiency and Welfare}

Without rankings-dependent utility, RSD always produces an assignment that is Pareto efficient ex-post: the first agent in the randomly chosen RSD ordering, say $i_1$, is assigned to her top-ranked object. The second agent, say $i_2$, is assigned to her highest-ranked object that was not taken by $i_1$, so the only way to make $i_2$ better off is to give her the object that went to $i_1$ (assuming $i_2$ prefers it), but this makes $i_1$ worse off. Continuing this argument, it is easy to see that each successive agent can only be made better off by taking the assignment of some earlier agent, and thus at least one of these agents must be worse off. 

However, once agents' fundamental values and rankings-dependent utilities are taken into account (as opposed to just ordinal preferences), RSD may no longer perform as well on welfare grounds. For instance, consider 3 agents and 3 goods, with the following fundamental values. 
\begin{center}
    
\begin{tabular}{c|ccc}
 & $v_{i}(x)$ & $v_{i}(y)$ & $v_{i}(z)$\tabularnewline
\hline 
$i_{1}$ & 1 & 0.7 & 0\tabularnewline
$i_{2}$ & 1 & 0.7 & 0\tabularnewline
$i_{3}$ & 1 & 0.7 & 0\tabularnewline
\end{tabular}

\end{center}
Also, let $\rho(1)=0.1$ for all agents, and $\rho(2)=\rho(3)=0$. It is easy to check that truth-telling with respect to fundamental values (i.e., all agents report ordinal preferences $xPyPz$) is the unique equilibrium of RSD so the issue in the previous section is not relevant. No matter the random ordering, all final allocations result in a sum of utilities that is equal to $W^{RSD}=(1+0.1)+0.7+0=1.9$.

Next, consider Boston with the following strategies: agents $i_1$ and $i_2$ report truthfully with respect to fundamental values, while agent $i_3$ reports $yPzPx$. It can be checked that this is an equilibrium.  In this equilibrium, one of the agents---agent $i_3$ in our example---reports her second-best good, good $y$, first. This ensures that she gets $y$ for sure, but also leaves the remaining two agents with a higher chance of receiving the best good, $x$ (in particular, agents $i_1$ and $i_2$ each have a 50/50 chance of receiving $x$ and $z$). Since the preferences of $i_1$ and $i_2$ are symmetric, any final allocation results in a sum of utilities $W^B=(1+0.1)+(0.7+0.1)+0=2 > W^{SD}$. 

Thus, Boston results in a greater overall total welfare. Intuitively, the reason is that Boston incentivizes some agents to rank the good with the second-highest fundamental value first. In equilibrium, this means that there will be two agents who are getting a good ranked first, and so, both of these agents will receive the rankings-dependent utility $\rho(1)$. This is our motivation for comparing the RSD and Boston mechanisms in our experiment.  In the next section, we will provide formal theorems and corresponding experimental hypotheses that generalize these intuitions for our experimental environment.\footnote{There is a recent strand of literature that also emphasizes that non-strategyproof mechanisms may outperform strategyproof ones in equilibrium even with just standard preferences because the opportunity to ``misrepresent'' their preferences gives a channel by which agents can express some information on their cardinal utilities \citep{abdul_che_yasuda:AER:2011,Troyan:2012,abdulkadirouglu2015expanding,fragiadakis2019designing}. While this channel is still present in our model, additionally including rankings-dependent utility will amplify the welfare gains of non-strategyproof mechanisms.}

\section{Experimental Design}\label{sec:Experimental-Design}

The experiment consists of two phases that participants complete in succession.  We have two treatments that only differ in the matching mechanism run in Phase II.  These treatments are \emph{RSD} which runs the random serial dictatorship and \emph{Boston} which runs the Boston mechanism.  Hereafter, we will continue to use italics for the treatment names and regular text for the mechanisms.  We will describe each phase here, and the full instructions are provided in Appendix D.

\subsection{The Two Phase Experiment}

\noindent{\it Phase I}

Phase I is identical for both treatments.  The participants complete 22 incentivized tasks.  The first 20 tasks consist of valuing 20 different objects with the multiple price list elicitation method.  See Appendix C for a description of each of the 20 objects including the monetary values (measured as the prices on Amazon.com where we purchased the objects) and average elicited values.

The multiple price lists are framed as willingness to accept (WTA) and consist of two screens. The first screen elicits the value of each object in dollar increments and the second screen elicits the value in two cent increments.  In particular, for each of the 20 objects, the participants are told they have been allocated the object and then have the opportunity to exchange the object for various amounts of money.  On the first screen, they see a list where each row represents keeping the object or exchanging it for an amount of money that ranges from  \$1.00 to \$50.00.  The participants choose the row that is the last row where they prefer keeping the object to exchanging it.  We provide a screen shot in Figure \ref{fig:PhaseI1} in Appendix D where the participant has selected the row with the dollar amount \$16.00.  The participants can change their minds and click a different row, and then confirm their choices once they have come to a final decision.  The second screen is presented identically to the first except that the amounts of money range from \$$x$.02 to \$$x$+1 in \$0.02 increments where $x$ is the value of the last row (the one they had selected) from the first screen.  We provide a  screen shot in Figure \ref{fig:PhaseI2} in Appendix D where the participant has selected the dollar amount \$16.56.

In order to assist the participants in this valuation task, pictures of each object are provided on the screens as the objects are valued.  We also had the physical objects at the front of the room and would bring them over for further inspection on request.  All participants value the frisbee first, the set of picture frames second, the cable spirals third, and the final 17 objects in random order.  The reason to fix the first three objects is to allow participants to gain familiarity with the elicitation method.  These three objects are not relevant for Phase II and are not analyzed in the results section.

If one of these valuation tasks is selected for payment for a given participant, we pay the participant in the standard way for multiple price lists.  We randomly and uniformly draw a number between 1 and 50 which corresponds to a dollar amount the participant will receive.  The participant then receives the object if they had indicated that they preferred the object to that amount of money and the dollar amount if they had indicated that they preferred the amount of money to the object.  In the case that the randomly drawn number is exactly equal to the dollar amount in the last row where they would keep the object (the row they had clicked on the first screen), then a second randomly and uniformly drawn number between 1 and 50 is generated.  In this case, the number corresponds to the two cent increments on the second screen (1 was \$0.02, 2 was \$0.04, etc.) and the participant receives the object if they had indicated that they preferred the object to the amount of money and the dollar and cents if they had indicated the opposite.

The final two tasks are incentivized risk aversion and loss aversion elicitation tasks.  The risk aversion task is the classic \cite{holt2002risk} price list task.  To be consistent with the previous 20 tasks, there are 50 rows of lottery choices between Lottery A, high payoff of \$24.00 with $x\%$ chance and low payoff of \$20.00 with $100-x\%$ chance, and Lottery B, high payoff of \$38.00 with $x\%$ chance and low payoff of \$12.00 with $100-x\%$ chance.  The chance of the high payoff, $x$, increases across rows from 2\% to 100\%.  The participants select the last row where they preferred Lottery A.  If this task is selected for payment for a given participant, we randomly and uniformly draw a number between 1 and 50 and run Lottery A if that number is less than or equal to the number of the last row where they preferred Lottery A, and we run Lottery B otherwise.

The loss aversion task compares risky choices with gains to risky choices with losses.  Again, there are 50 rows of lottery choices.  For this task, the participant chooses between Option A, \$20.00 for sure plus a 50\% chance of a \$10.00 bonus, and Option B, \$30.00 for sure plus a 50\% chance of losing \$x.  The loss in Option B varies from \$20.00 to \$0 in increments of \$0.40 across the rows.  The participants select the last row where they prefer Option A.  If this task is selected for payment for a given participant, we randomly and uniformly draw a number between 1 and 50 and run Option A if that number is less than or equal to the number of the last row where they preferred Option A, and we run Option B otherwise.

Finally, the participants complete an unincentivized questionnaire to conclude the phase.  The questionnaire includes the standard cognitive reflection task questions as well as demographics and academic endeavors.
\bigskip

\noindent{\it Phase II}

After completing the 22 tasks and questionnaire in Phase I, the participants move on to Phase II.  In this phase, the participants are randomly matched into groups of five to engage in a matching market with the five goods. We use the same five goods for Phase II in every session, and they are all taken from the set of twenty goods valued in Phase I.  The goods for Phase II are:
\begin{enumerate}
    \item Fjallraven backpack: A small 16L gray backpack from the popular brand Fjallraven with a monetary value of \$66.95.\footnote{The goods were purchased on Amazon.com.  The prices on Amazon vary slightly from day to day so the monetary values given here are approximate.}
    \item Hydroflask waterbottle: A light-blue reusable water bottle from the popular brand Hydroflask with a monetary value of \$32.96.
    \item Moleskine notebook: A notebook with 192 pages and a high-quality black cover from the popular brand Moleskine with a monetary value of \$21.90.
    \item Blue ceramic mug: A generic blue ceramic mug with a monetary value of \$11.99.
    \item Set of 4 Uni-ball pens: A set of four fine-point black rollerball pens from the well-known brand Uni-ball with a monetary value of \$6.88.
\end{enumerate}

Each participant engaged in a matching market corresponding to their treatment, \textit{RSD} or \textit{Boston}, just one time.  They then value the good they received from the mechanism using the exact same multiple price list elicitation method used for the 20 goods valued in Phase I.

At the end of the experiment, participants were paid for a single task that was equally likely to be one of the tasks in Phase I or the unique task in Phase II.  In the case a task from Phase I was selected, each of the 22 tasks was equally likely to be selected for payment.  As there are more tasks in Phase I, each is less likely to be selected for payment than the Phase II task. We discuss the reasons for this payment structure in Remark \ref{rem:endowment-effect} in the next subsection.

\begin{rem}
\normalfont The participants only engage in a single matching market.  We chose this method so that they would feel greater ownership of the good they received.  In order to make sure the participants fully understood the procedure we presented them with instructions that included an example (with different goods than in the actual market), had them complete a quiz regarding a second example which required them to answer all the questions correctly before moving on, and provided them with an 8 minute practice period in which they could engage in as many markets as they wanted against robot players.  We feel comfortable that they understood the mechanisms, because they almost all received 100\% on the quiz the first time, rarely used the full 8 minutes to practice against the robots, and because neither the Boston nor the RSD mechanisms are particularly complicated mechanisms.  
\end{rem}

\subsection{Theoretical Predictions and Experimental Hypotheses}

\subsubsection*{Rankings-Dependent Utility}

Our first and main hypothesis assesses whether preferences are indeed rankings-dependent.

Recall our model with rankings-dependent utility, Equation (\ref{eq:rank-dep-utility-function}):
\begin{equation*}
u_i(x,P_i) = v_i(x) + \rho(j).     
\end{equation*}
To test the hypothesis that preferences are rankings-dependent, we are interested in the term $\rho(j)$. Our experimental design allows us to recover exactly this.  Suppose that, in the matching mechanism in Phase II, a participant $i$ receives object $x$, which was reported as her $j^{th}$ ranked good.  The valuation elicitation at the end of Phase II measures $v_i(x) + \rho(j)$. In Phase I, the participant also valued object $x$, independent of any mechanism or ranking context. Thus, the Phase I valuation of object $x$ is $v_i(x)$.  So the net value, $NV(j)$, the difference in valuations for the same object $x$ between Phase II and Phase I, is the rankings-dependent component $\rho(j)$:
\begin{equation}\label{eq:Net-Value}
NV(j)=\text{Phase II value - Phase I value}=(v_i(x)+\rho(j))-v_i(x)=\rho(j)
\end{equation}
If $\rho(j)$ is decreasing in $j$ as assumed in Section \ref{sec:Experimental-Design}, the same should be true for the net value between Phase I and Phase II. This gives rise to our first hypothesis.\footnote{Several commenters have mentioned that we might also want to consider the percentage change in Net Value---i.e., calculate Net Value = (Phase II value - Phase I Value)/(Phase I value)---in addition to a linear formulation.  A problem arises in that this does not allow us to cleanly disentangle the fundamental value from rankings dependence. This is because we would have $NV(j)=\rho(j)/v_i(x)$, and we would be unable to determine whether any changes in NV were due to changes in $\rho$ or changes in $v$.}

\begin{hyp}
Participants prefer to get objects they rank higher, and so $NV(j)$ will be
decreasing in submitted rank $j$.
\end{hyp}

\begin{rem} \normalfont
One might be concerned that in the course of the experiment subjects are asked to value one object twice, and a desire to appear ``consistent'' could bias the results. First, notice that such a desire actually works against finding an effect, as it will tend to compress Net Value towards 0; in other words, if we do find an effect, it will not be driven by the desire for consistency. Second, while impossible to eliminate completely, we designed the experiment to mitigate this concern as much as possible. By placing the questionnaire and explanation of the matching mechanisms in the middle of the experiment (so that, by the time they submitted their Phase II valuation, a lot of time had passed) and having them value 20 objects in Phase I (15 of which were not relevant for the experiment), we are confident that most subjects did not remember their initial valuation for the good that they received in Phase II.
\end{rem}

\begin{rem}\label{rem:endowment-effect}[Endowment Effect]
\normalfont The ``endowment effect" is the well-established finding that participants value objects more when they own them than when they do not \citep{marzilli2014endowment}. One of the leading explanations for the endowment effect is expectations-based loss aversion. Expectations-based loss aversion is also a potential explanation for rankings-dependent utility in matching mechanisms, as shown by \cite{dreyfuss2022deferred} (cf. the Introduction, where we discuss this in more detail). However, there are also two other sources of ``endowment effects'' that could be present in our experiment: the elicitation mechanism itself and the probability that a given task is selected for payment.

To the extent that rankings-dependent is due to expectations-based loss aversion, the first potential endowment effect falls under what we are trying to measure, while the other sources are not of interest to us. With this in mind, we designed the experiment specifically so that these latter two phenomena do not affect our hypotheses. First, we use the same WTA elicitation mechanism in both Phases so any endowment effect from the elicitation mechanism immediately cancels out when calculating Net Value.  The probability of payment effect is a little trickier,  but it is also not an issue for testing our hypotheses because they are statements about comparisons across Net Values for which an added constant is irrelevant.

To make this point about probability of payment more formally, we can augment Equation \ref{eq:rank-dep-utility-function} by adding a ``probability of payment" endowment effect term to the utility function as follows:
$$
u_i(x,P_i)=v_i(x)+\rho(j)+\xi_k
$$
where $\xi_k$ represents the endowment effect for phase $k=I,II$. Because the Phase II task has a higher probability of being chosen for payment, we expect $\xi_{II} > \xi_{I}$. Then, Net Value (equation \ref{eq:Net-Value}) becomes 
\begin{eqnarray*}
NV(j)&=&\text{Phase II value - Phase I value}\\
&=& (v_i(x)+\rho(j)+\xi_{II})-(v_i(x)+\xi_I)\\
&=&\rho(j)+\xi_{II}-\xi_I.
\end{eqnarray*}
Hence, it is true that Net Value entangles rankings-dependence with the probability-of-payment endowment effect so we cannot make any statements about the exact values we elicit for Net Value. Importantly, though, when we compare across rankings, the probability-of-payment endowment effect constant is irrelevant. Specifically, Hypothesis 1 predicts that Net Value is decreasing across rank which is still the case when we add a constant to all Net Values. 

In other words, one possible explanation for rankings-dependence is a difference in endowment effects between receiving a good from a mechanism in which it is ranked and receiving it independent of any mechanism context, and our experiment is designed to pick this up. However, our agnosticism on the source of rankings dependence allows for other explanations as well. The goal of our experimental design was to provide direct evidence of rankings dependence that might arise from many sources, rather than look at one specific microfoundation.

\end{rem}

\subsubsection*{Truth-telling}

Our next two hypotheses concern truth-telling and preference-reporting. Recall that, in the standard matching model without rankings-dependent utility, RSD is strategyproof while the Boston mechanism is not.  In the case of Boston, agents may manipulate their preferences implied by their fundamental values by ranking popular objects lower, because they are likely to be taken in earlier rounds of the mechanism. But how do these properties extend to the case of rankings-dependent utility?

Truth-telling with respect to fundamental values will no longer be a dominant strategy of either mechanism, because agents may want to manipulate and rank objects that are less popular highly so that they receive a higher-ranked object and the corresponding increase in $\rho(j)$. In general, it is difficult to solve for an equilibrium with rankings-dependence. However, by putting more structure on the model, we will be able to show some results. In particular, we make the following assumption: 
\medskip

\noindent{\bf Assumption:} For all $x_j\in X$, $v_i(x_j)=v_{i'}(x_j):=v_j$ for all $i,i'\in I$ and that these values are common knowledge. Further, $v_1,v_2 \gg \bar{v}=v_3=\cdots=v_n$.
\medskip

In words, we assume that there is complete information, a common ordinal preference, and that objects $x_1$ and $x_2$ are much better than the other objects.\footnote{Though unlikely to ever hold exactly, common ordinal preferences is an assumption commonly made to approximate highly correlated preferences while still maintaining analytic tractability (see, e.g., \cite{abdul_che_yasuda:AER:2011}, \cite{featherstone2016boston}, and \cite{fragiadakis2019designing}). As we explain in this paragraph, we think our preference assumption captures well the fundamental strategic tension we wanted to induce with our selection of objects, which is whether to rank the backpack or the water bottle first.} The motivation for this assumption is that goods 1 and 2 are strongly enough preferred by all agents to all other goods, so that the main strategic decision is whether to rank $x_1$ or $x_2$ first. The exact goods in the experiment were chosen with this model of valuations in mind: given the popularity of Fjallraven backpacks and Hydroflask water bottles to our participant pool of undergraduate students, we expected these two objects to be the most popular, and strongly preferred to the other three objects (a notebook, coffee mug, and pens). Indeed, 84\% of the participants value the backpack and water bottle as the two best goods, as measured by the Phase I elicitation. Table \ref{phase-I-goods} in the appendix shows that the average elicited values for the five objects used in the matching mechanism were \$28.24 (backpack), \$22.56 (water bottle), \$9.11 (notebook), \$6.53 (mug) and \$5.33 (pens).

 Given the preference assumption, for the theory, we focus on equilibria that have the following structure. For RSD: $n_1$ of the agents rank $x_1$ first and $x_2$ second; the remaining $n-n_1$ agents rank $x_2$ first and $x_1$ second; for the remaining goods $x_3,\ldots,x_n$, each agent draws a ranking of these goods uniformly at random from all possible rankings. For Boston: $n_1$ agents rank $x_1$ first; the remaining $n-n_1$ agents rank $x_2$ first;  for the remaining goods $x_3,\ldots,x_n$, as in RSD, each agent draws a ranking of these goods uniformly at random from all possible rankings. They rank these goods immediately below their top choice, and rank the remaining good that was not their top choice (either $x_1$ or $x_2$) at the bottom of their list.
 
 In other words, in RSD, all agents rank $x_1$ and $x_2$ first and second, in some order. In Boston, each agent ranks either $x_1$ or $x_2$ first, and puts the other option not chosen last in their preferences. This is in line with our motivation that the main source of competition is over goods $x_1$ and $x_2$. Thus, an agent's strategy boils down to whether to rank $x_1$ or $x_2$ first, and the key equilibrium object to solve for is the exact number of agents that choose to rank $x_1$ (the best object according to fundamental value) first in their reported preferences in RSD versus Boston. As we show in the proof of Theorem \ref{thm-first-gooods-first}, this number is uniquely determined for each mechanism.\footnote{\label{fn:boston-caveat}There is one caveat to the strategies defined above, which is that if all agents rank $x_1$ first in Boston ($n_1=n$), then, given the preference assumption, it is optimal for everyone to rank $x_2$ second, rather than last, in their preference list. Essentially, this happens when $v_1\gg v_2$ to the extent that it is worth it for every agent to enter the round 1 lottery for $x_1$, in which case $x_2$ will still be available in round 2 of the mechanism, and Boston effectively becomes equivalent to RSD. The theorems stated below still hold regardless. See the proofs in Appendix \ref{sec:proofs-of-theorems} for details.}     

\begin{thm}
\label{thmtt}
     Suppose the preference assumption holds. If truth-telling with respect to the fundamental value is an equilibrium of the Boston mechanism, truth-telling is also an equilibrium of RSD.
\end{thm}

This theorem suggests that there should be more truth-telling in RSD compared to Boston. Intuitively, truth-telling is an equilibrium of the Boston mechanism when the fundamental value $v_1$ is so high that it is worth it for every agent to enter the round 1 lottery for $x_1$, rather than deviating by ranking $x_2$ first and receiving payoff $v_2+\rho(1)$ for sure. In RSD, this deviation is even less profitable, because it does not guarantee $x_2$ with certainty (though it does guarantee the deviating agent will not get $x_1$). Thus, if the deviation is not profitable in Boston, it will not be profitable in RSD, either, and so truth-telling with respect to fundamental values is an equilibrium of both mechanisms. 

Our experimental design allows us to measure each participant's rankings with respect to fundamental value, because the fundamental values are elicited in Phase I. This gives our second testable hypothesis.

\begin{hyp}
Participants will submit preferences in Phase II that are truthful with respect to fundamental values as measured by the elicitation in Phase I weakly more in RSD than in Boston.
\end{hyp}

Theorem \ref{thmtt} only discusses the truth-telling equilibrium. It may of course be that truth-telling is not an equilibrium of either Boston or RSD: Once rankings-dependent utility is introduced, even in RSD, an agent may want to deviate from truthful reporting in order to obtain the additional rankings-dependent utility term $\rho(1)$. Even in this case, however, we still expect more agents to misreport by ranking $x_2$ first in the Boston mechanism, for similar reasons as for Theorem \ref{thmtt}. This is formalized in the next theorem.

\begin{thm}\label{thm-first-gooods-first}
Suppose the preference assumption holds. Then, in any equilibrium of RSD, weakly more agents rank good $x_1$ first in their preference list than in any equilibrium of Boston.
\end{thm}

This theorem leads to our third hypothesis.

\begin{hyp}
    In Phase II, weakly more participants will rank the good with the highest elicited Phase I value first in their preference list in RSD than in Boston.
\end{hyp}

\subsubsection*{Welfare}

Our final hypothesis concerns the overall welfare of the two mechanisms. Arguably, this is the most important feature of any mechanism, because the ultimate goal of any allocation mechanism is to produce an outcome that maximizes participant satisfaction, taking into account all components of utility, including fundamental values and rankings-dependent components.

\begin{thm}
\label{thmwel}
 Suppose the preference assumption holds. The equilibrium total welfare of the Boston mechanism is weakly higher than the equilibrium total welfare of RSD.
\end{thm}

The intuition for this result is that, given our preference assumption, the sum of the fundamental value components of total welfare is the same for any allocation. Thus, the welfare comparison between any two mechanisms is determined by the sum of the rankings-dependent utility terms:
$$
W=\sum_{j=1}^n (\#\text{ agents who receive their $j^{th}$ ranked good})\times \rho(j)
$$
In the Boston mechanism, in equilibrium, at least one agent ranks good $x_2$ first and receives it for sure, and so there are two agents who receive rankings-dependent utility $\rho(1)$.\footnote{There is also the case that $v_1$ is so high that all agents rank $x_1$ first in Boston. However, in this case, all agents will also rank $x_1$ first in RSD (see Theorem \ref{thm-first-gooods-first}) and the mechanisms become equivalent, so the welfare of the two mechanisms will be the same.} In RSD, even if some agents do rank $x_2$ first, there is still a non-zero probability that the good will go to an agent who ranked it second, and who thus will receiving rankings-dependent utility $\rho(2)<\rho(1)$, resulting in lower total welfare for RSD. The full details of the argument can be found in the appendix. 

In the experiment, we measure the total welfare of a mechanism as the sum of the elicited values for the mechanism allocation in Phase II. This gives our final hypothesis.

\begin{hyp}
    The sum of the elicited values in Phase II will be weakly higher in Boston than in RSD. 
\end{hyp}
Each of the theorems and experimental hypotheses above compare RSD and Boston using weak comparisons. For instance, they suggest that total welfare under Boston will be weakly higher than under RSD. Although there are valuation vectors that make each comparison strict, for others, the outcomes will be equal. A limitation of our experimental design is that we cannot control the valuation vectors as is possible with an induced value design. For this reason, we consider Hypothesis 1—the one that doesn't rely on any such assumptions about the $v$'s—as the primary point of interest. At the same time, we still think Theorems \ref{thmtt}-\ref{thmwel} and their corresponding hypotheses are valuable for providing novel insights into the effects of rankings-dependent preferences on incentives and welfare, and so include them in our results as well.

\subsection{Procedures}
All experiments were run at the University of Virginia VeconLab with undergraduate students recruited from the Darden BRAD lab recruitment pool.  There were a total of 200 participants, 100 for each of the two treatments in a between-subjects design. No feedback was provided before the end of the experiment so each participant is treated as an independent observation for the results.  The experiment was programmed and run with z-Tree \citep{fischbacher2007z}.  For the participants who earned money the average earnings were \$33.59.  For the participants who received goods, the average monetary values of these goods was \$36.69.\footnote{In total, 144 participants earned money and 56 participants received goods.} Additionally, all participants received \$6 for showing up.

\section{Results}\label{sec:results}
In this section, we turn to the data to analyze the extent to which our experimental results provide support for our four hypotheses.  For regressions we use stars to indicate significance at the usual levels ($*$ if $p<.1$, $**$ if $p<.05$, and $***$ if $p<.01$).

\subsubsection*{Hypothesis 1: Net Value}
\begin{figure}[t]
  \centering
  \includegraphics[width=.8\linewidth]{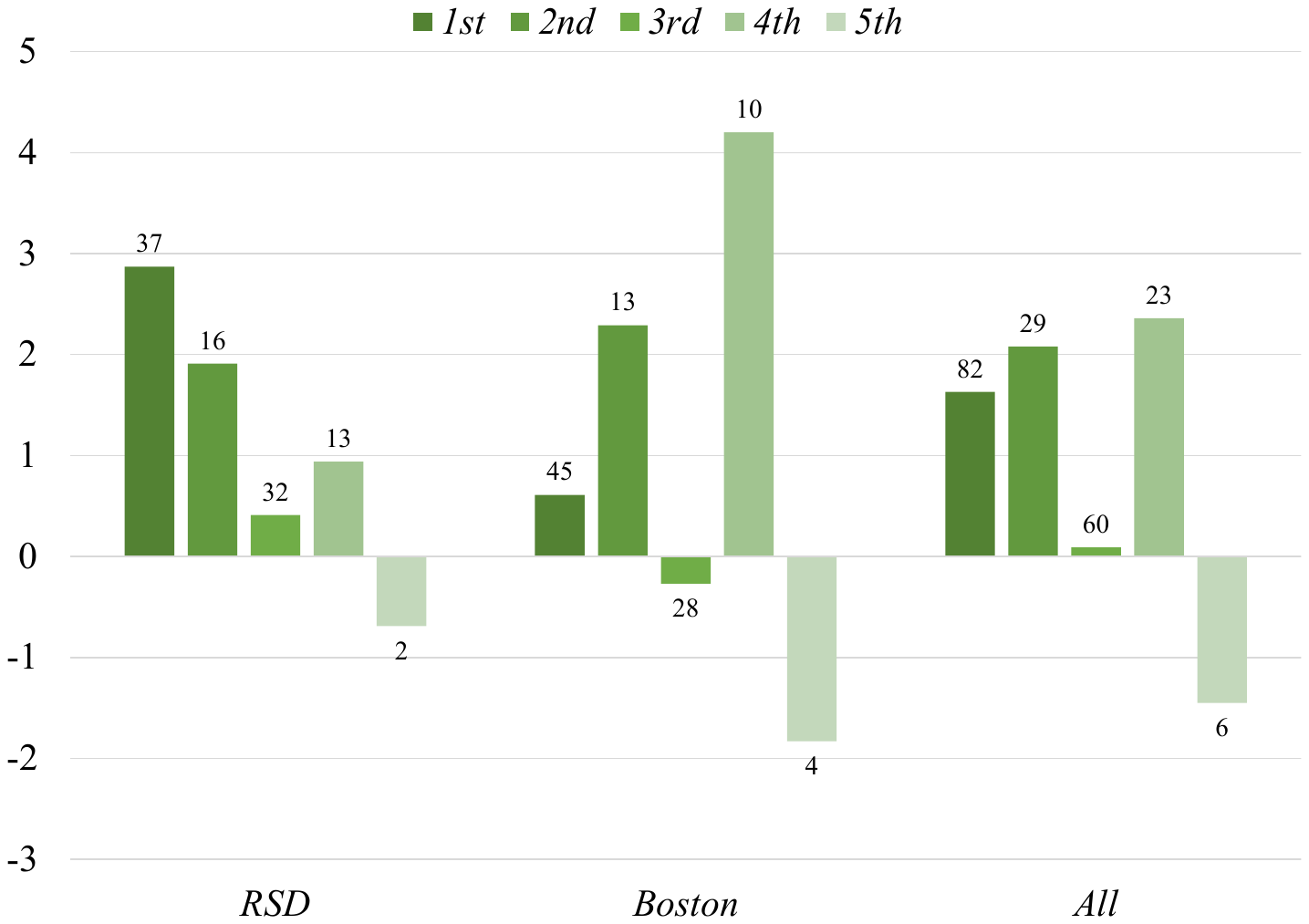}  
  \caption{Net Values with \# of obs. on ends of bars}
  \label{fig:NV}
\end{figure}

Recall that Hypothesis 1 argues that net value, the elicited value in Phase II minus the elicited value in Phase I, will be decreasing in submitted rank $j$.  As a first look, Figure \ref{fig:NV} provides the average net values for each rank for $RSD$, $Boston$, and both treatments.

For \emph{RSD}, the results presented in Figure \ref{fig:NV} are consistent with the hypothesis.  Net value is almost monotonically decreasing, with the exception of the $3^{rd}$ and $4^{th}$ rankings ($n$ is only 13 for the latter case, as most people received one of their top 3 ranked goods).  Indeed, a non-parametric test rejects the null of no difference among the ranks in favor of the alternative that net value is decreasing in submitted rank (Jonckheere-Terpstra, $p=0.019$).  On the other hand Figure \ref{fig:NV} seems to show no relationship between net value and submitted rank in  \emph{Boston}, and the non-parametric test does not reject the null in this case (Jonckheere-Terpstra, $p=0.618$).

To explore these results in more detail, we turn to regression analysis.  The regressions are presented in Table 1. For each treatment, we first regress net value on submitted rank alone (regressions (1) and (3)).  We then add a number of potential influences on net value including Phase I elicited value (initial value), Holt-Laury switching point (risk aversion), the loss aversion task switching point (loss aversion), score out of three on the three cognitive reflection tasks (CRT score), a dummy for gender (female), the randomly selected order in which the good received in Phase II was valued in Phase I (Phase I order), the number of submitted preference orders in the practice session against robots (practice), and a dummy for whether the received good in Phase II was ranked higher than it would have been according to the implied fundamental values elicited in Phase I (truthful).  These regressions are reported in columns (2) and (4), and the pooled regression with both treatments is presented in column (5).

\begin{center}
\input{regressionsFormat.tex}
\end{center}

In line with the non-parametric tests, Table 1 provides support for the hypothesis that net value is decreasing in rank of received good for \emph{RSD}.  Net value decreases by \$0.88 or \$1.24 per ranking depending on whether controls are included or not, and the estimates are marginally significant.  Surprisingly, in contrast to the non-parametric tests, Table 1 also provides support for the hypothesis for \emph{Boston} when controls are included.  For \emph{Boston}, net value decreases by \$1.81 per ranking when controls are included and the estimate is significant.  Pooling the data, there is a strongly significant decline of \$1.55 per rank.

Additionally, in \emph{Boston}, participants' initial values negatively impact net value while risk aversion positively impacts net value.  The effect from initial value must be carefully evaluated, because initial value is part of the net value.  Observe that the estimate $-0.2768$ means that an increase of \$1.00 in initial value decreases net value by \$0.28, but this can also be translated and interpreted as an increase of \$1.00 in initial value increases final value (the Phase II valuation) by \$0.72.  Combined with the positive intercept in the regression, the estimates imply that small initial values are increased and large initial values are decreased.  One interpretation of the result then is reversion to the mean; participants perhaps value objects with noise and so low Phase I valuations are more likely to be increased while high Phase I valuations are more likely to be decreased.\footnote{For example, a small initial value is more likely to come from a participant who under-valued the good in Phase I relative to their true fundamental value and is therefore more likely to increase the valuation in Phase II.}


The effect of risk aversion is harder to understand.  As we will discuss in the next section, risk aversion impacts truth-telling in \emph{Boston}.  Briefly, we show there that risk averse participants are more likely to highly rank a good that they do not like as much but is less popular.  Here, this means that when they get this less popular good their net values increase.  We are not as confident as to why this may be the case, but one possibility is utility from the relief that they do indeed obtain the good they moved up in their rankings.


Given the discrepancy between the raw data and the regression analysis (particularly with the difficult to interpret positive effect of risk aversion in \emph{Boston}), we think it is prudent to dive a little deeper.  As a check of robustness, we ran additional regressions with dummy variables for final rank and for the good received.  The first set of dummies allows for a non-linear relationship between rank and net value, while the second allows for the possibility that there is something inherent about the good itself.  In order to not have too many regressors relative to data, we only included the controls for initial value and risk aversion.  The results are presented in Table 5 in the appendix.

These results show some similarities and some differences to Table 1. Most notably, the support for the hypothesis goes away.  In particular, in \emph{RSD}, the estimates that are significant are for receiving the highest-monetary goods, the backpack and the waterbottle.\footnote{The mug is omitted.  So the coefficients are interpreted with respect to the mug which has the smallest net value.}  The results for \emph{Boston} indicate it is only initial value and risk aversion that matter.  Putting them together, there is no evidence in support of the hypothesis in Table 5.  So is Table 1 or Table 5 correct?  Table 5 has higher values of $R^2$, but we also do not want to say it is more correct than Table 1.  Indeed, the reason we have it as a check of robustness is that there are two reasons to be wary of the results in Table 5.  First, the dummy variables reduce the power of the estimates for the effects of rank (essentially only considering 20 data points at a time rather than the whole set of 100).  Indeed the estimates for ranks in \emph{Boston} are negative indicating that rank does decrease net value, just not significantly, which could be an issue of power.  Second, there is clearly high collinearity between the goods received, the initial value, and the final ranks.  For example, Table 1 indicates that getting your $1^{st}$ ranked good is best for net value while Table 5 indicates that getting the backpack is best for net value.  Of course, these two outcomes are highly correlated (and backpacks also had the largest initial values so that is correlated as well).  This is the reason we omit the goods dummies in our main regressions reported in Table 1.  

To conclude, we find mixed evidence for Hypothesis 1: the evidence seems stronger for  \emph{RSD} than for \emph{Boston}, which may indicate that the choice of mechanism may influence preference formation in other more complex ways as well. This should be further explored in the future with alternative experimental designs. Part of the ambiguity may be due to the use of real goods, which may induce more noise in the data, making strong statistical inference more difficult. This is a necessary trade-off to answer our question of interest, because real goods allow us to directly identify rankings-dependence in ways that induced values would not. 

\subsubsection*{Hypotheses 2 and 3: Truth-telling}

We now move to assessing truth-telling (with respect to fundamental value) and Hypotheses 2 and 3.  To assess Hypothesis 2, we begin with reporting the proportion of participants whose submitted rankings list in Phase II is truthful with respect to the fundamental values elicited in Phase I.\footnote{The preferences elicited in Phase I are not always strict, in which case listing the indifferent goods in any order is classified as truthful.} 

When testing Hypothesis 2, we classify an agent as truth-telling only if all five goods were ranked in the same order as the fundamental value elicited in Phase I. However, this could be an overly restrictive definition of truth-telling, because the goods were intentionally chosen such that two of them (the backpack and water bottle) were likely to be much better than the other three to most people. To the extent that the remaining three goods (notebook, mug, pen) were viewed as significantly worse than the top two, and similar in value amongst themselves, participants may have been more focused on the choice between the top two goods and submitted ``noisy'' preferences over the remaining three goods. There also is likely to be noise in the Phase I elicitation procedure itself, and since the notebook, mug, and pen were similar (and lower) in value, this could be a cause of apparently non-truthful reporting over all 5 goods in Phase II.\footnote{Indeed, the interpretation of the initial value coefficient in Table 1 suggests that this is the case.} 

We deal with this issue in two ways. First, we also assess Hypothesis 3, which looks at the number of agents who report their top good truthfully. Second, we consider submitted rankings lists in Phase II that are truthful with respect to the fundamental values elicited in Phase I with some noise.  In particular, we chose an ad-hoc cutoff of \$2 and classify preferences as truthful as long as they do not rank a good in Phase II higher than another good with an elicited value that is \$2 more.  All the proportions are presented in Table 2.

\begin{center}
\begin{tabular}{lccc}
\multicolumn{4}{c}{Table 2: Truth-telling Rates}\\ \hline
Measure & \emph{RSD} & \emph{Boston}& All \\ \hline
Exact & & & \\
\hspace{2mm} All &0.40 &0.37 &0.39 \\
\hspace{2mm} Top Choice &0.78  &0.79&0.79 \\
\hspace{2mm} Top 2 Choices &0.60 &0.60  &0.60 \\ \hline
Up to \$2 differences & & & \\
\hspace{2mm} All &0.58 &0.53 &0.56 \\
\hspace{2mm} Top Choice &0.82  &0.83&0.83 \\
\hspace{2mm} Top 2 Choices &0.70 &0.67  &0.69 \\ \hline

\end{tabular}
\end{center}

Of course, as the leniency allowed in truthful reporting increases, the proportions of truthful reporting increases.  But the main result, regardless of how truth-telling is measured, is that there are no differences in truthful reporting between the two treatments.  The non-parametric Wilcoxon ranksum test confirms this impression for all 6 pairwise comparisons between the treatments.  We find no support that Hypotheses 2 and 3 hold strictly.  We also find it interesting that even in the most lenient case, about 20\% of participants are putting a good at the top of their submitted list that they valued more than \$2 less than their top-value good. 

To provide a clearer picture of what impacts truth-telling, we ran probit regressions with a dependent variable equal to 1 for participants who chose the truth and the independent variables risk aversion, loss aversion, CRT score, female, and practice (all measured the same as in the net value regressions).  We report the results for the case where truth-telling is measured as the top 2 choices in the main text, because we think that misstating the bottom 3 goods is mostly noise and it is exactly these two goods for which there is an incentive to misrepresent one's preferences in the Boston mechanism.  The other two measures are presented in Tables 6 and 7 in the Appendix where we find estimate values are quite similar although the levels of significance vary somewhat.

\begin{center}
\input{regressionstttFormat.tex}
\end{center}

The regressions indicate that while truth-telling rates are similar across the two treatments, this may be due to different reasons.  The estimate for CRT score is positive and marginally significant in \emph{RSD} (with the exact measure, regression (1)) indicating that deeper-thinking participants tell the truth more often.  So perhaps the non truth-telling in \emph{RSD} is due to mistakes made by participants with low CRT scores rather than driven by rankings-dependent utility.  The estimate for risk aversion is negative and marginally significant in \emph{Boston} (with the exact measure, regression (3)) indicating that risk averse participants are less likely to tell the truth, which is line with a model of standard preferences. Finally, the estimate for female is positive and significant in \emph{Boston} which is consistent with the finding that females are more averse to lying \citep{dreber2008gender,erat2012white}.

\subsubsection*{Hypothesis 4: Welfare}

Finally, we address Hypothesis 4 and welfare.  Table 4 presents the results for welfare measured as the sum of participants elicited Phase II values for the goods they received in the matching mechanism.

\begin{center}
\begin{tabular}{ccc}
\multicolumn{3}{c}{Table 4: Welfare}\\ \hline
\emph{RSD} & \emph{Boston} & All \\ \hline
90.29 & 89.97 & 90.13\\
\hline 
\end{tabular}
\end{center}

As is clear in the table, there is no difference between the two treatments.  This is confirmed with the Wilcoxon rank sum non-parametric test.\footnote{It may also be argued that rankings-dependent utility is a ``bias'' that should not be included in the final welfare comparisons (we thank an anonymous referee for pointing this out). In this case, a better measure of welfare would be the sum of the Phase I valuations. Under this measure, the average welfare is 82.25 for RSD and 85.75 for Boston. So, there is slightly higher welfare for Boston, but the difference is not significant.}  
\section{Conclusion}

We investigate whether agent preferences over goods received via a matching market are influenced by how highly they ranked the object in their reported preference list. We design a laboratory experiment to test whether  agents value a good more the higher that they rank it. A novel feature of our experiment is that we use real goods, rather than induced values. This more faithfully simulates real-world settings in which participants must form their own values over their potential outcomes, and allows us to directly measure for rankings-dependent utility by eliciting values for objects both inside and outside of the mechanism and taking the difference. 

We find mixed evidence to support the hypothesis that valuations are influenced by reported rankings, with a clearer effect seen for RSD than for Boston. We also find much lower rates of truth-telling in RSD than previous matching experiments that use induced values designs, which is consistent with preferences being rankings-dependent. While we do not claim a ``slam-dunk'' for the hypothesis of rankings dependence, given their potential implications, we think the results of our real-goods experiment warrant further study of the issue of preference formation in matching mechanisms, which has been largely ignored in the (vast) matching  literature thus far. This will probably require the use of more real-goods experimental designs as opposed to induced values. While induced values have the advantage of giving the experimenter more control, by construction, they are unable to say anything about preference formation. Using real goods will induce new challenges in experimental design, but with a significant benefit of more closely replicating and understanding the challenges faced by people in real-world matching markets.  

\newpage

\appendix
\section{Proofs of Theorems}\label{sec:proofs-of-theorems}

\textbf{Proof of Theorem \ref{thmtt}.} By supposition of the theorem, assume that truth-telling is an equilibrium of the Boston mechanism. We will show that truth-telling is also an equilibrium of RSD by showing that if all $j\neq i$ report truthfully, then $i$'s best response is also to report truthfully. 

 Let all agents other than $j\neq i$ report a preference ranking that respects the fundamental values, $P_j:x_1,\ldots,x_n$. Fixing these reports for $j\neq i$, let $i$'s expected utility from any report $P_i'$ be $EU^{SD}(P_i')$. If $i$ also reports a ranking that respects the fundamental values, $P_i:x_1,\ldots,x_n$, this becomes
\begin{equation}\label{SD-truthful-payoff}
EU^{SD}_i(P_i)=\frac{1}{n}\sum_{j=1}^n u_i(x_j,P_i)=\frac{1}{n}\sum_{j=1}^n (v_j+\rho(j)).
\end{equation}
Consider any alternative report for $i$, $P_i'\neq P_i$. Let $k'$ be the lowest index such that object $x_{k'}$ is not ranked in the $k'$-th position, and let $x_{k^*}$ be the object that is ranked $k'$-th, where by definition, $k^*>k'$. Notice that such a $k^*$ exists, as otherwise $P_i'=P_i$. Under $P_i'$, agent $i$ will never receive any object $x_{k'},\ldots,x_{k^*-1}$. To see this, let $\ell$ be $i$'s (randomly drawn) order in the serial dictatorship. If $\ell<k'$, then $i$ will receive object $x_\ell$; if $k'\leq \ell \leq k^*$, object $x_{k^*}$ is still available when it is $i$'s turn to choose, and thus $i$ receives object $k^*$; if $\ell > k^*$, then at $i$'s turn, all objects $x_1,\ldots, x_{k^*}$ have been taken by earlier agents, and thus $i$ receives some object from the set ${x_{k^*+1},\ldots,x_n}$. This implies that $i$'s utiliy from reporting $P_i'$ is bounded above by 
\begin{equation}\label{SD-deviation-payoff}
EU^{SD}(P_i')\leq \frac{1}{n}\sum_{j=1}^{k'-1}(v_j+\rho(j))+\frac{n-k'+1}{n}(v_{k^*}+\rho(k'))
\end{equation}
Now, consider the Boston mechanism, and the same report $P_i'$ from above (continuing to assume that all other agents report truthfully in Boston). If $i$ reports $P_i'$ in the Boston mechanism, then her expected utility is
\begin{equation}\label{B-deviation-payoff}
EU^{B}(P_i')=\frac{1}{n}\sum_{j=1}^{k'-1}(v_j+\rho(j))+\frac{n-k'+1}{n}(v_{k^*})+\rho(k'))
\end{equation}
This is because, in Boston, if $i$ is not assigned in one of the first $k'-1$ rounds, she is assigned to $x_{k^*}$ in round $k'$ with certainty. In this case, she has ranked this object $k'$-th, so her total utility is the term in parentheses at the end of equation (\ref{B-deviation-payoff}). This event occurs with probability $(n-k'+1)/n$. 

Notice that if $i$ reports truthfully with respect to fundamental values in Boston, her payoff is 
\begin{equation}\label{B-truthful-payoff}
EU^{B}_i(P_i)=\frac{1}{n}\sum_{j=1}^n (v_j+\rho(j)),
\end{equation}
which is equivalent to the payoff from reporting $P_i$ in SD; see equation (\ref{SD-truthful-payoff}). Finally, we have the following:
\begin{eqnarray}\label{SD-truthtell-optimal}
    EU^{SD}(P_i')\leq EU^{B}(P_i')\leq EU^{B}(P_i)=EU^{SD}(P_i) 
\end{eqnarray}
where the first inequality follows from equations (\ref{SD-deviation-payoff}) and (\ref{B-deviation-payoff}), the second from the fact that truth-telling is assumed to be an equilibrium of Boston, and the last equality from equations (\ref{SD-truthful-payoff}) and (\ref{B-truthful-payoff}). Equation (\ref{SD-truthtell-optimal}) implies $EU^{SD}(P_i')\leq EU^{SD}(P_i)$, i.e., truth-telling is optimal in SD. $\hfill \blacksquare$ 

\vspace{5mm}

\textbf{Proof of Theorem \ref{thm-first-gooods-first}.} First, consider Boston, and let $n_1^B$ be the number of agents who rank $x_1$ first. It is trivial to see that $n_1^B=0$ is never an equilibrium, because if no agent is ranking $x_1$ first, then any agent can deviate to a strategy that does so and receive payoff $v_1+\rho(1)$ for sure, which is the highest possible attainable payoff. 

Next, consider the case that $n_1^B=n$ is an equilibrium of the Boston mechanism, i.e., all agents rank $x_1$ first. In this case, each agent will also rank $x_2$ second (see footnote \ref{fn:boston-caveat}), and so receives $x_1$ with probability $1/n$ and $x_2$ with probability $1/n$, for an equilibrium utility of $(1/n)\times(v_1+\rho(1))+1/n\times(v_2+\rho(2))+((n-2)/n)\times (\bar{v}+\bar{\rho})$, where $\bar{\rho}=\frac{1}{n-2}\sum_{j=3}^n\rho(j)$. The last term, $\bar{v}+\bar{\rho}$, comes from the fact that for all agents not assigned in rounds 1 or 2, the assignment is just a random assignment of goods $x_3,\ldots,x_n$, and so by symmetry, any individual agent's payoff is just the average, $\bar{v}+\bar{\rho}$; this event happens with probability $(n-2)/n$.   If any individual agent deviates to ranking $x_2$ first, she gets it for sure, with resulting utility $v_2+\rho(1)$. Since this is an equilibrium, we conclude that $(1/n)\times(v_1+\rho(1))+1/n\times(v_2+\rho(2))+((n-2)/n)\times (\bar{v}+\bar{\rho}) \geq v_2+\rho(1)$. Note also if $n_1^B=n$ is an equilibrium, then it is the unique equilibrium. This follows because if there were some other equilibrium in which $n_1^B<n$, then some agent $i$ is not ranking $x_1$ first. Agent $i$ will therefore never receive $x_1$, and so her payoff is bounded above by by $v_2+\rho(1)$. If she deviates to ranking $x_1$ first and $x_2$ second, then her payoff is bounded below by $(1/n)\times(v_1+\rho(1))+1/n\times(v_2+\rho(2))+((n-2)/n)\times(\bar{v}+\bar{\rho})$. As was just shown, the latter is greater than the former, and so this deviation is profitable.

Now, consider SD. If all $n$ agents rank $x_1$ first and $x_2$ second, then they once again receive payoff $(1/n)\times(v_1+\rho(1))+1/n\times(v_2+\rho(2))+((n-2)/n)\times (\bar{v}+\bar{\rho})$. If any agent deviates to ranking anything other than $x_1$ first,\footnote{This follows because if the agent is chosen first in the RSD ordering, she receives her first-ranked object which is different from $x_1$, while if she is not chosen first, then whoever is chosen first takes $x_1$.} they will never receive $x_1$, and so their payoff is bounded above by $v_2+\rho(1)$, which, from the above calculation for Boston, is smaller than the expected payoff from ranking $x_1$ first and $x_2$ second. Thus, $n_1^{SD}=n$ is also an equilibrium of SD. Similar arguments as for the Boston case above show that this equilibrium is once again unique, and so $n_1^{SD}\geq n_1^{B}$, as required.

So, for the remainder of the proof, we restrict to $0<n_1<n$. Let $U^B(x_j,n_1)$ be the expected utility of an agent who ranks good $x_j$ first when $n_1$ total agents (including this agent) rank $x_1$ first, and the remaining $n-n_1$ total agents rank $x_2$ first. Then, we can calculate:
\begin{align*}
    U^B(x_1,n_1)&=\frac{1}{n_1}(v_1+\rho(1))+ \frac{n_1-1}{n_1}\delta \\
    U^B(x_2,n_1)&= \frac{1}{n-n_1}(v_2+\rho(1)) + \frac{n-n_1-1}{n-n_1}\delta 
\end{align*}
where $\delta=\bar{v}+(\rho(2)+\rho(3)+\ldots+\rho(n-1))/(n-2)$.\footnote{This is slightly different than $\bar{v}+\bar{\rho}$ calculated above for the $n_1=n$ case, because here, agents begin ranking goods $x_3,\ldots,x_n$ second in their list, and so the summation of the rankings-dependent utility terms run from $j=2$ to $n-1$; they will never receive their $n^{th}$ ranked object, which is the object among $\{x_1,x_2\}$ that they did not rank first.} The equations derive from the fact that if an agent ranks $x_1$ first in Boston, she has a $1/n_1$ chance of getting $x_1$. If she does not, good $x_2$ will not be available in round 2, and so she will get some good $x_3,\ldots,x_n$. There will always be exactly $n-2$ agents left after round 1 of the mechanism, and since all agents rank goods $x_3,\ldots,x_n$ the same, this becomes a random allocation of the remaining objects.  The $\delta$ terms represent the total expected utility from this random allocation. Let 
$$\Delta^B_{x_1\rightarrow x_2}(n_1)=U^{B}(x_1,n_1) - U^{B}(x_2,n_1-1)$$
be the change in utility for an agent whose equilibrium strategy is to rank $x_1$ first, and who deviates to ranking $x_2$ first. Similarly, let 
$$\Delta^B_{x_2\rightarrow x_1}(n_1)=U^{B}(x_2,n_1) - U^{B}(x_1,n_1+1)$$
be the expected change for an agent whose equilibrium strategy is to rank $x_2$ first, but who deviates to ranking $x_1$ first. 

For $n_1^B$ to be an equilibrium, we need both $\Delta^B_{x_1\rightarrow x_2}(n_1^B)\geq 0$ and $\Delta^B_{x_2\rightarrow x_1}(n_1^B)\geq 0$, i.e., no agent has a profitable deviation.
Algebra shows that these two equations reduce to
\begin{equation}\label{boston-range}
n_1^B\in \left[ \frac{n(v_1+\rho(1)-\delta)}{v_1+v_2+2(\rho(1)-\delta)}-\frac{v_2+\rho(1)-\delta}{v_1+v_2+2(\rho(1)-\delta)}, \frac{n(v_1+\rho(1)-\delta)}{v_1+v_2+2(\rho(1)-\delta)}+\frac{v_1+\rho(1)-\delta}{v_1+v_2+2(\rho(1)-\delta)} \right].
\end{equation}
 Subtracting the left endpoint from the right endpoint gives a range of length exactly 1. So, there will be a unique integer $n_1^B$ in this range, which corresponds to the unique equilibrium number of agents who rank $x_1$ first in Boston.

Let $n_1^{SD}$ be the equilibrium number of agents who rank $x_1$ first in SD. We will show that $n_1^{SD}\geq n_1^B$. For SD, the analogous equations to the above are: 
\begin{align*}
    U^{SD}(x_1,n_1)&=\frac{1}{n}(v_1+\rho(1)) + \frac{1}{n}\left(\frac{n_1-1}{n-1}(v_2+\rho(2))+\frac{n-n_1}{n-1}(v_1+\rho(1))\right) + \frac{n-2}{n}\delta'\\ \bigskip
    U^{SD}(x_2,n_1)&= \frac{1}{n}(v_2+\rho(1)) + \frac{1}{n}\left(\frac{n_1}{n-1}(v_2+\rho(1))+\frac{n-n_1-1}{n-1}(v_1+\rho(2))\right)+\frac{n-2}{n}\delta'
\end{align*}
In each equation above, there is a $1/n$ chance that the agent is ordered first, in which case she gets her first ranked good. If not, there is a $1/n$ chance she is ordered second. In the top equation, there is an $(n_1-1)/(n-1)$ chance the first agent was one of the remaining agents who ranked $x_1$ first, in which case agent $i$ receives $x_2$, and a $n_2/(n-1)$ chance the first agent was one of the agents who ranked $x_2$ first, in which case $i$ gets $x_1$ again. If $i$ is ordered third or higher in the SD ordering, then both goods $x_1$ and $x_2$ are gone at her turn, and, as for the Boston case above, $i$'s utility in this case is that from a random assignment of the remaining goods, represented by $\delta'=\bar{v}+\bar{\rho}$.\footnote{The term $\delta'$ here is slightly different from the $\delta$ term in the Boston equations above, because in RSD, agents rank $x_1$ and $x_2$ first and second, in some order, while in Boston, one of $x_1$ or $x_2$ is ranked last. However, for RSD, the $\delta'$ terms are less important, because they will cancel out when checking for profitable deviations below.} 

Define $\Delta^{SD}_{x_1\rightarrow x_2}(n_1)$ and $\Delta^{SD}_{x_2\rightarrow x_1}(n_1)$ analogously to the above, except replacing SD for Boston. Similarly as for Boston, for $n_1^{SD}$ to be an equilibrium, we need $\Delta^{SD}_{x_1\rightarrow x_2}(n_1^{SD}), \Delta^{SD}_{x_2\rightarrow x_1}(n_1^{SD})\geq 0$. Algebra shows that these equations reduce to
\begin{equation}\label{sd-range}
n_1^{SD} \in \left[\alpha + \frac{1}{2}, \alpha -\frac{1}{2} \right]
\end{equation}
where $\alpha= \frac{n}{2}+\frac{(n-1)}{2}\frac{v_1-v_2}{\rho(1)-\rho(2)}$. Once again, this range has total length 1, and so there is a unique equilibrium number of agents $n_1^{SD}$ that rank $x_1$ first. 

What remains to check is that $n_1^{SD}\geq n_1^B$. To show this, note first that the lower bounds in equations (\ref{boston-range}) and (\ref{sd-range}) are determined by the equations $\Delta^B_{x_2\rightarrow x_1}(n_1)\geq 0$ and $\Delta^{SD}_{x_2\rightarrow x_1}(n_1)\geq 0$ that ensure that those who rank $x_2$ first do not want to deviate to ranking $x_1$. Next, notice that
$$
\frac{d\Delta^{SD}_{x_2\rightarrow x_1}(n_1)}{dn_1}=\frac{2(\rho(1)-\rho(2))}{n(n-1)}>0,
$$
which implies that $\Delta^{SD}_{x_2 \rightarrow x_1}(n_1)$ is an increasing function. Now, evaluate $\Delta^{SD}_{x_2 \rightarrow x_1}(n_1)$ at $\xi=\frac{n(v_1+\rho(1)-\delta)}{v_1+v_2+2(\rho(1)-\delta)}-\frac{v_2+\rho(1)-\delta}{v_1+v_2+2(\rho(1)-\delta)}$, which is the lower bound of equation (\ref{boston-range}):
$$
\Delta^{SD}_{x_2 \rightarrow x_1}\left(\xi\right)=-\frac{(v_1-v_2)((n-3)\rho(1)+(n+1)\rho(2)+(n-1)(v_1+v_2)-2(n-1)\delta)}{n(n-1)(v_1+v_2+2\rho(1)-2\delta)}
$$
Now, since $\bar{v}<v_2$ and $\rho(j)$ is a decreasing function, we have 
$$
\delta=\bar{v}+\frac{1}{n-2}\sum_{j=2}^{n-1}\rho(j)<v_2+\rho(2)
$$
It is then simple to check that both the numerator and denominator of the above equation are positive, which implies that $\Delta^{SD}_{x_2 \rightarrow x_1}\left(\xi\right)<0$. Because the function is increasing, the crossover point that defines the lower bound of equation (\ref{sd-range}) must lie to the right of $\xi$. Thus, the range of equation (\ref{sd-range}) must be to the right of the range of equation (\ref{boston-range}) (they may overlap), which implies that $n_1^{SD}\geq n_1^B$. $\hfill \blacksquare$

\medskip

\textbf{Proof of Theorem \ref{thmwel}.} The overall welfare from any mechanism is just the sum total of the utilities. Recall that an agent's utility is the sum of the fundamental value of the object she receives, $v_j$, and a rankings-dependent component: $u_i(x,P_i)=v_i(x)+\rho(j)$. Because $v_i(x_j)=v_j$ for all $i$, the sum of the fundamental values will be the same for any assignment. This means that the overall welfare of any mechanism $\psi$ is determined by the sum of the rankings-dependent utility components:
$$
W^{\psi}=\sum_{j=1}^n (\# \text{ agents who receive their $j^{th}$ ranked good}) \times \rho(j)
$$
There are two cases. First, if $n_1^B=n$---that is, all agents choose to rank $x_1$ first in the Boston mechanism---then, by Theorem \ref{thm-first-gooods-first}, $n_1^{SD}=n$ as well. The two mechanisms are then equivalent, and so $W^{B}=W^{SD}$. 

Second, assume that $1\leq n_1^B <n$ (recall that Theorem \ref{thm-first-gooods-first} shows that $n_1^B=0$ is never an equilibrium of Boston). For Boston, we have 
$$
W^B = 2\rho(1)+\sum_{j=2}^{n-1} \rho(j).
$$
The first term, $2\rho(1)$ comes from the fact that at least one agent is ranking $x_1$ first and at least one agent is ranking $x_2$ first, and so both of these agents are receiving their top-ranked object. Then, all remaining agents rank the remaining goods uniformly at random, and so, by symmetry, they are equally likely to get any of these goods. Notice that the summation starts at 2, because they place the good from $\{x_1,x_2\}$ that was not their top choice at the bottom of their rankings (and will never receive it).

For SD, the equation is
\begin{multline}
    W^{SD}=\frac{n_1^{SD}}{n}\frac{n_1^{SD}-1}{n-1}(\rho(1)+\rho(2))+2\times\frac{n_1^{SD}}{n}\frac{n-n_1^{SD}}{n-1}(\rho(1)+\rho(1)) \\ 
     +\frac{n-n_1^{SD}}{n}\frac{n-n_1^{SD}-1}{n-1}(\rho(1)+\rho(2))+\sum_{j=3}^n \rho(j)
\end{multline}

The first three terms derive from the probability that the first two agents are ``$x_1P_ix_2$'' agents or ``$x_2P_ix_1$'' agents. For instance, for the first term, there is an $n_1^{SD}/n$ chance that the first agent in the order ranks $x_1 P_i x_2$ and, conditional on this, a $(n_1^{SD}-1)/(n-1)$ chance that the second agent has the same ranking. In this case, $x_1$ goes to an agent who ranked it first and $x_2$ goes to an agent who ranked it second, so the rankings-dependent component of welfare is $\rho(1)+\rho(2)$. The next two terms are calculated similarly, for all possible combinations of the preferences of the first two agents. Finally, the summation term at the end comes from the fact that all remaining agents rank $x_3,\ldots,x_n$ uniformly randomly, and so by symmetry, are equally likely to get any rank. Note that, unlike in Boston, the summation runs from $j=3$ to $n$, because all agents rank $x_1$ and $x_2$ first in SD. 

Now, notice that
\begin{align*}
    W^{SD} &\leq  \frac{n_1^{SD}}{n}\frac{n_1^{SD}-1}{n-1}(2\rho(1))+2\times\frac{n_1^{SD}}{n}\frac{n-n_1^{SD}}{n-1}(2\rho(1))+\frac{n-n_1^{SD}}{n}\frac{n-n_1^{SD}-1}{n-1}(2\rho(1))+\sum_{j=3}^n \rho(j) \\
     &=2\rho(1)\left(\frac{n_1^{SD}}{n}\frac{n_1^{SD}-1}{n-1}+2\times\frac{n_1^{SD}}{n}\frac{n-n_1^{SD}}{n-1}+\frac{n-n_1^{SD}}{n}\frac{n-n_1^{SD}-1}{n-1} \right)+\sum_{j=3}^n \rho(j)\\
     &=2\rho(1)+\sum_{j=3}^n \rho(j)\\ 
     &\leq W^B
\end{align*}
where the first line replaces $\rho(2)$ with $\rho(1)\geq \rho(2)$, the second factors out the $2\rho(1)$ terms, the third follows because the term in parentheses sums to 1, and the last follows from $\rho(j)$ being a decreasing function. $\hfill \blacksquare$

\newpage 
\section{Robustness Regressions}

\begin{center}
\input{regressionsdumFormat.tex}
\end{center}

\begin{center}
\input{regressionsttFormat.tex}
\end{center}

\begin{center}
\input{regressionsttoFormat.tex}
\end{center}

\newpage
\section{Phase I goods}\label{sec:Phase-I-goods}

\begin{center}
\begin{table}[h]
\caption{Phase I goods}
\label{phase-I-goods}
\begin{tabular}{llcc}
\hline\hline
Good & Description & Amazon Price & Avg. Elic. Val.\\ \hline
{\bf Backpack} & {\bf Fjallraven gray, 16L backpack} & {\bf \$66.95}& {\bf \$28.24}\\
Alarm clock & Aisuo bluetooth alarm/speaker/nightlight & \$37.99 & \$19.46\\
{\bf Water bottle} & {\bf Hydroflask blue, 32 oz. water bottle} & {\bf \$32.96} & {\bf \$22.56}\\
Shower speaker & Donerton bluetooth waterproof speaker & \$29.99 & \$18.75 \\
Laptop stand & Ergonomic universal laptop stand & \$26.99 & \$15.99\\
Outdoor blanket & Bearz blue waterproof blanket & \$24.99 & \$14.04\\
Cold brewer & Takeya 1 qt. carafe & \$21.00 & \$15.04\\
Picture frames & Set of 4 white, wooden frames & \$20.99 & \$9.99\\
Popcorn set & 3 bags of popcorn plus flavors & \$22.00 & \$8.72\\
{\bf Notebook} & {\bf Moleskine 192 age, black notebook} & {\bf \$21.90} & {\bf \$9.11}\\
Tile & Bluetooth chip to track item on phone & \$19.99 & \$14.98\\
Phone mount & Gooseneck phone holder with bracket & \$19.79 & \$8.87\\
Popcorn popper & Silicone microwaveable popcorn maker & \$14.99 & \$7.49\\
Charging pad & Anker phone-charging pad & \$13.99 & \$14.57\\
Frisbee & Discraft yellow 175 g. Frisbee  & \$13.76 & \$6.60\\
{\bf Mug} & {\bf Blue ceramic mug} & {\bf \$11.99} & {\bf \$6.53}\\
Playing cards & Deck of black playing cards & \$7.99 & \$4.93\\
{\bf Pens} & {\bf Set of 4 Uni-ball rollerball pens} & {\bf \$6.88} & {\bf \$5.33}\\
Keychain tool & Multi-tool that attaches to keys & \$6.64 & \$5.11\\
Cable spirals & Set of 24 plastic, multicolor cable-protectors & \$6.29 & \$5.03\\ \hline
\end{tabular}
\end{table}
\end{center}
\newpage
\section{Instructions and Screen Shots}
\subsubsection*{Welcome and Phase I}
Welcome.  This is an experiment in decision making.  Various research foundations and institutions have provided funding for this experiment and you will have the opportunity to make a considerable amount of money which will be paid to you at the end.  Make sure you pay close attention to the instructions because the choices you make will influence the amount of money you will take home with you today.  Please ask questions if any instructions are unclear.

{\bf Terminology:} In the following instructions, we will say that the computer will make a {\bf random} choice from a number of possibilities. This means that the computer will randomly select one of the possibilities with equal chance for each. If there are N possibilities, you can think of this as the computer rolling a die that has N sides, and choosing the possibility that comes up on the die. 

This experiment will consist of two phases.  We will hand out the instructions for each phase before you complete the phase.  You will be paid for your choices in only one of the two phases which will be randomly selected by the computer.  Your choices in each phase have no impact on later phases.

Your earnings may be an object or money, and at the end of the experiment, we will give you the object or the money to take home. You will see pictures of the object on your screen, but we have the actual objects here with us.  If you want us to show you the actual object at any time, just raise your hand and we can bring it over.  Everyone will also get \$6 for participating.

{\bf Phase I}

For this phase, you will be allocated an object that will be shown to you on your screen.  Remember that we have the actual objects with us, so feel free to raise your hand if you would like us to bring one over for closer inspection.

{\bf The Task}

After you are allocated your object, you will have the opportunity to give up the object in exchange for a certain amount of money. Whether you keep the object or receive money will be determined as follows:

First, you will see a screen with 50 rows.  Each row is a choice between keeping the object or exchanging it for \$1, \$2, …, up to \$50  . As we will explain carefully in the earnings section of the instructions below, if a row is selected for your earnings, you will take home the choice (object or money) that you selected in that row. Because it is time-consuming to have you click a button for every row, instead you only need to {\bf click on the bullet for the object in the last row where you would keep the object over the specified number of dollars.} Because all rows above the one you select offer less money, we will fill in all of these rows with you selecting to keep the object. Similarly, because all rows below offer more money, we will fill in all of these rows with you selecting the money.  After you click, you can change your mind by clicking on a different row and that row will become the last row where you keep the object.  Click confirm when you have finalized your choice.

Next, you will see another screen with 50 rows.  Each row is a choice between keeping the object or exchanging it for \$x.02, \$x.04, up to \$x+1 where x is the dollar amount in the row you selected on the first screen.  As for the first screen, you make just one choice.  {\bf Click on the bullet for the object in the last row where you would keep the object over the specified number of dollars.} As on the first screen, because all rows above have less money, we will automatically fill in all of these rows with you selecting the object over the money. Because all rows below contain more money, we will automatically fill in all these rows with you selecting the money over the object. After you click, you can change your mind by clicking on a different row and that row will become the last row where you keep the object. Click confirm when you have finalized your choice.

{\bf Procedures}

You will do this task for {\bf 20 objects}.  After you complete the task for the 20 objects, we will ask you to complete 2 additional tasks unrelated to this task for a total of 22 tasks in Phase I.  The instructions for those tasks will be presented on the screen when you do them.  Finally, we will have you answer a short questionnaire.

{\bf Earnings}

If this phase is selected for earnings, the computer will first randomly select 1 of the 22 tasks and you will receive earnings for that task.  If the selection is one of the 20 tasks described above, your earnings will be determined as follows.  The computer will randomly select one of the 50 rows from the first screen.  {\bf We will then implement your choice from that row.  That is, you will get the object to take home with you if you selected to keep the object in this row, or the amount of money if you selected the money.}

There is one exception:  If the randomly selected row from the first screen is the last row where you would keep the object, the computer will randomly select one of the 50 rows from the second screen and {\bf we will then implement your choice from that row.} Keep in mind that any row from any task could be selected so it is in your best interest to select the object when you prefer the object to the money and to select the money when you prefer the money to the object.  

For the 2 additional tasks, how your earnings would be determined if the computer selected one of them will be explained to you when you complete those tasks .
\newpage
\subsubsection*{Phase II: RSD}

{\bf Phase II}

For this phase, you will be divided into groups of 5 people, and for each group there will be 5 objects. Each member of the group will end up with exactly one of the objects. The object you get will be determined by your preferences for the objects, along with the preferences of your group members. Each group member will provide a {\bf preference ranking} of the objects. Because there is only one of each object available, if multiple people list a particular object first, we will use a random tie-breaking procedure to determine who gets what. The details of how this is done are described below. 

{\bf The Task}

You will first be shown the 5 objects. Your task for this phase is to {\bf submit a list ranking the objects from most-preferred to least-preferred.}
After collecting everyone’s rankings, the computer will randomly select a picking ordering of the 5 group members. The computer will then allocate an object to each group member as follows.
\begin{itemize}
\item	The group member who is randomly chosen to go first will be allocated the first object on their submitted list.
\item	The group member who is randomly chosen to go second will be allocated the highest-ranked object on their list that was not taken by the first group member. 
\item	The group member who is randomly chosen to go third will be allocated the highest-ranked object on their list that was not taken by the first or second group member. 
\item	The group member who is randomly chosen to go fourth will be allocated the highest-ranked object on their list that was not taken by the first, second, or third group member. 
\item	The group member who is randomly chosen to go fifth will be allocated the highest-ranked object on their list that was not taken by the first, second, third, or fourth group member. 
\end{itemize}

It is up to you how to rank the objects.   Given the rules of the procedure, there is no way to manipulate your rankings to obtain an object that is ranked higher than the one you would have ended up with if you had just ranked the objects in the order you prefer them.  This is because your ranking list does not affect what objects will be available when it is your turn in the picking order to receive an object, and when your turn arrives, the computer will give you the object you have ranked highest among those that are still available.

{\bf Example:}

Let’s go through an example.  For our example we will use four objects that are not used in the experiment and a group of four people to make it simpler to understand and ensure we are making no suggestions about how you should rank the objects in the experiment.  The example is purely to help you understand the procedure the computer will use.

We will use the names Ann, Bob, Carol, and Dave. The items are a Pizza, Chips, Soda, and Pretzels. Suppose our group members submit the following rankings:
\begin{center}
    \begin{tabular}{|c|c|c|c|c|}\hline
    	&Ann	&Bob&	 Carol&	Dave\\ \hline
1st choice &	Chips	&Pizza&	Pizza&	Pretzels\\ \hline
2nd choice&	Pizza&	Chips&	Pretzels&	Soda\\ \hline
3rd choice&	Soda&	Soda&	Chips&	Chips\\ \hline
4th choice	&Pretzels&	Pretzels&	Soda&	Pizza\\ \hline
    \end{tabular}
\end{center}

After submitting their rankings, the computer randomly determines a picking ordering of the group members. Say that the computer randomly orders Ann, Bob, Carol, and Dave as follows: 
\begin{center}
    \begin{tabular}{|c|c|}	\hline
1st	&Bob\\ \hline
2nd&	Carol\\ \hline
3rd &	Dave\\ \hline
4th &	Ann\\ \hline
    \end{tabular}
\end{center}

The computer then determines the allocations using this ordering and the preferences of each group members using the following procedure
\begin{enumerate}
    \item 
	Bob was ordered first. His top-ranked object is the Pizza. Thus, he is given the Pizza. This indicated in {\bf bold} in the table.
		\begin{center}
    \begin{tabular}{|c|c|} \hline
    &Bob \\ \hline

1st choice&	{\bf Pizza}\\ \hline
2nd choice&	Chips\\ \hline
3rd choice&	Soda\\ \hline
4th choice&	Pretzels\\ \hline
    \end{tabular}
\end{center}
 \item	Carol was ordered next. The Pizza is gone, so the objects that remain are the Soda, the Chips, and the Pretzels. According to Carol’s list, her top choice of these is the Pretzels, so Carol is given the Pretzels. 
 	\begin{center}
    \begin{tabular}{|c|c|} \hline
	 &Carol\\ \hline
1st choice&	\sout{Pizza}\\ \hline
2nd choice&	{\bf Pretzels}\\ \hline
3rd choice&	Chips\\ \hline
4th choice&	Soda\\ \hline
    \end{tabular}
\end{center}
 \item	Dave was ordered third. The Pizza and the Pretzels are gone, so the objects that remain are the Soda and the Chips. According to Dave’s list, his top choice between the Soda and the Chips is the Soda, so he is given the Soda.
 	\begin{center}
    \begin{tabular}{|c|c|} \hline
	&Dave\\ \hline
1st choice&	\sout{Pretzels}\\ \hline
2nd choice&	{\bf Soda}\\ \hline
3rd choice&	Chips\\ \hline
4th choice&	\sout{Pizza}\\ \hline
     \end{tabular}
\end{center}
 \item	Ann is ordered last. The Pizza, Pretzels, and Soda are gone, so Ann receives the Chips.
 	\begin{center}
    \begin{tabular}{|c|c|} \hline
	& Ann\\ \hline
1st choice&	{\bf Chips}\\ \hline
2nd choice&	\sout{Pizza}\\ \hline
3rd choice&	\sout{Soda}\\ \hline
4th choice&	\sout{Pretzels}\\ \hline
    \end{tabular}
\end{center}
\end{enumerate}

{\bf Procedures:}

During the procedure, all you must do is submit {\bf one list ranking all of the objects.} Note that you will not know your place in the picking order when you submit your rankings. The computer will take everyone’s list and then determine the picking order. The picking order is entirely random, and is not influenced by the list you submit. As described above, when your turn comes, the computer will consider all of the objects that are still left, and give you the one that you ranked the highest.

After you are allocated your object, {\bf we will ask whether you want to keep the object or exchange it for various amounts of money using the exact same procedures as in Phase I.}

{\bf Earnings:}

If this phase is selected for payment the computer will randomly select rows just as in Phase I to determine if you will keep the object allocated to you by the procedure or the amount of money. 

{\bf Practice:}

We will now hand out a worksheet with another example of the procedure.  Please work through the worksheet and raise your hand when you are finished.  We will come over and check your work and help you if there are any mistakes.  

Then, there will be an 8 minute practice period on the computer.  The practice period will allow you to practice allocating 5 objects to 5 people.  You can submit a ranking list and the computer has robots who submit the other 4 ranking lists. While you will not know the random ordering in the actual procedure when you submit your rank list, this practice will allow you to simulate what you would have gotten if you had submitted different lists. You can experiment with different rank lists as many times as you like for the randomly selected picking order. You can also generate a new random picking order, and experiment further. 

For the practice round, we recommend that you first think about the order in which you would actually prefer the objects, and take note of the outcome you get for each possible list you try.  

We are doing this practice because you will only submit your ranking list once for the actual experiment and we want to make sure everyone fully understands the procedure before we continue to the actual experiment.  You do not earn anything for the practice.
\newpage
\subsubsection*{Phase II: Boston}

{\bf Phase II}

For this phase, you will be divided into groups of 5 people, and for each group there will be 5 objects. Each member of the group will end up with exactly one of the objects. The object you get will be determined by your preferences for the object, along with the preference of your group members. Each group member will provide a {\bf preference ranking} of the objects. Because there is only one of each object available, if multiple people list a particular object first, we will use a random tie-breaking procedure to determine who gets what. The details of how this is done are described below. 

{\bf The Task}

You will first be shown the 5 objects. Your task for this phase is to {\bf submit a list ranking the objects from most-preferred to least-preferred.}

After collecting everyone’s rankings, the computer will randomly assign each group member a number from 1 to 5 that will be used to break ties. Every group member will be assigned a different number.  The computer will then allocate an object to each group member in rounds as follows.
\begin{itemize}
  \item 	In the {\bf first round}, the computer looks at everyone’s {\bf first choices.}
  \begin{itemize}
   \item 	If only one person has an object as their first choice, that person is allocated the object.
   \item 	If more than one person lists an object as their first choice, then the computer will give the object to the person who was assigned the \underline{lowest number} 1-5.
   \end{itemize}
   \item 	In the {\bf second round}, only people who did not receive anything in the first round participate. In the second round, the computer looks at everyone’s {\bf second choices.}
   \begin{itemize}
   \item 	If only one person has an object as their second choice, then that person is allocated the object.
   \item 	If more than one person has an object as their second choice, then the computer will give the object to the person who was assigned the \underline{lowest number} 1-5.
   \end{itemize}
   The procedure continues in rounds in the same manner, considering only agents who remain and their third choices in round 3, fourth choices in round 4, etc., until everyone is assigned an object.
\end{itemize}

It is up to you how to rank the objects. The random numbers 1-5 are not assigned until after everyone submits their rankings. They will only be used if the computer needs to break ties.

{\bf Example:}

Let’s go through an example.  For our example we will use four objects that are not used in the experiment to make it simpler to understand and ensure we are making no suggestions about how you should rank the objects in the experiment.  The example is purely to help you understand the procedure the computer will use.

We will use the names Ann, Bob, Carol, and Dave. The items are a Pizza, Chips, Soda, and Pretzels. Suppose our group members submit the following rankings:
\begin{center}
    \begin{tabular}{|c|c|c|c|c|}\hline
&	Ann	&Bob&	 Carol	&Dave \\ \hline
1st choice&	Pizza&	Pizza&	Pizza&	Pretzels \\ \hline
2nd choice&	Pretzels&	Chips&	Chips&	Soda \\ \hline
3rd choice&	Chips&	Soda&	Pretzels&	Chips \\ \hline
4th choice	&Soda&	Pretzels&	Soda&	Pizza \\ \hline
\end{tabular}
\end{center}

After collecting the rankings, the computer randomly assigns each group member a number, in this case 1 to 4 because there are 4 group members. Say that this random assignment resulted in the following:
\begin{center}
    \begin{tabular}{|c|c|}\hline
Name &	Random Number\\ \hline
Ann  &	2\\ \hline
Bob &	1\\ \hline
Carol & 	3\\ \hline
Dave  &	4\\ \hline
\end{tabular}
\end{center}

{\bf Round 1}
\begin{itemize}
\item	The computer considers the first choice of every group member (indicated in gray in the table below).
\item	Only Dave has the pretzels as his first choice, so Dave gets the pretzels.
\item	Ann, Bob, and Carol have the pizza as their first choice. 
\item	Bob has a lower random number (1) than Ann (2) or Carol (3). Bob gets the pizza.
\item	Ann and Carol get nothing in this round.
\item	The objects that are assigned are denoted in {\bf bold} in the table below. 
\end{itemize}
\begin{center}
    \begin{tabular}{|c|c|c|c|c|}\hline
&	Ann	&Bob&	 Carol	&Dave\\ \hline
\cellcolor{gray!50} 1st choice&	\cellcolor{gray!50}Pizza&\cellcolor{gray!50}	{\bf Pizza} &	\cellcolor{gray!50}Pizza& \cellcolor{gray!50}{\bf	Pretzels}\\ \hline
2nd choice&	Pretzels&	Chips&	Chips&	Soda\\ \hline
3rd choice&	Chips&	Soda&	Pretzels&	Chips\\ \hline
4th choice&	Soda&	Pretzels&	Soda&	Pizza\\ \hline
\end{tabular}
\end{center}

{\bf Round 2}
\begin{itemize}
\item	Only Ann and Carol participate in Round 2. The remaining objects are the Chips and the Soda. 
\item	The computer considers Ann and Carol’s second choices.
\item	Carol’s second choice is the chips. Carol receives the chips.
\item	Ann’s second choice is the pretzels, but the pretzels were already taken by Dave in round 1. 
\item	Ann does not receive anything in this round.
\end{itemize}
\begin{center}
    \begin{tabular}{|c|c|c|}\hline
&	Ann&	 Carol\\ \hline
1st choice&	\sout{Pizza}&	\sout{Pizza}\\ \hline
\cellcolor{gray!50}2nd choice&\cellcolor{gray!50}	\cellcolor{gray!50}\sout{Pretzels}&	\cellcolor{gray!50}{\bf Chips}\\ \hline
3rd choice&	Chips&	Pretzels\\ \hline
4th choice&	Soda&	Soda\\ \hline
\end{tabular}
\end{center}

{\bf Round 3}

\begin{itemize}
\item	Only Ann remains. The computer considers Ann’s third choice.
\item	Ann’s third choice is the chips, but the chips were already taken by Carol in Round 2.
\item	Ann receives nothing in this round.
\end{itemize}
\begin{center}
    \begin{tabular}{|c|c|}\hline
	&Ann\\ \hline
1st choice&	\sout{Pizza}\\ \hline
2nd choice&	\sout{Pretzels}\\ \hline
\cellcolor{gray!50}3rd choice&\cellcolor{gray!50}	\sout{Chips}\\ \hline
4th choice&	Soda\\ \hline
\end{tabular}
\end{center}

{\bf Round 4}

\begin{itemize}
\item	Only Ann remains. The computer considers Ann’s fourth choice.
\item	Ann’s fourth choice is the soda.
\item	Ann receives the soda.
\end{itemize}
\begin{center}
    \begin{tabular}{|c|c|}\hline
	&Ann\\ \hline
1st choice&	\sout{Pizza}\\ \hline
2nd choice&	\sout{Pretzels}\\ \hline
3rd choice&	\sout{Chips}\\ \hline
\cellcolor{gray!50}4th choice&\cellcolor{gray!50}	{\bf Soda}\\ \hline
\end{tabular}
\end{center}

{\bf Procedures:}

During the procedure, all you must do is submit {\bf one list ranking all of the objects.} The computer will take everyone’s list and determine assignments in each round based on the rank lists following the above procedure, breaking ties using the randomly assigned numbers.

After you are allocated your object, {\bf we will ask whether you want to keep the object or exchange it for various amounts of money using the exact same procedures as in Phase I.}

{\bf Earnings:}

If this phase is selected for payment the computer will randomly select rows just as in Phase I to determine if you will keep the object allocated to you by the procedure or the amount of money. 

{\bf Practice:}

We will now hand out a worksheet with another example of the procedure.  Please work through the worksheet and raise your hand when you are finished.  We will come over and check your work and help you if there are any mistakes.  

Then, there will be an 8 minute practice period on the computer.  The practice period will allow you to practice allocating 5 objects to 5 people.  You can submit a ranking list and the computer has robots who submit the other 4 ranking lists. While you will not know the random ordering in the actual procedure when you submit your rank list, this practice will allow you to simulate what you would have gotten if you had submitted different lists. You can experiment with different rank lists as many times as you like for the randomly selected picking order. You can also generate a new random picking order, and experiment further. 

For the practice round, we recommend that you first think about the order in which you would actually prefer the objects, and take note of the outcome you get for each possible list you try.  

We are doing this practice because you will only submit your ranking list once for the actual experiment and we want to make sure everyone fully understands the procedure before we continue to the actual experiment.  You do not earn anything for the practice.

\newpage
\begin{figure}[H]
\begin{subfigure}{\textwidth}
  \centering
  \includegraphics[width=\linewidth]{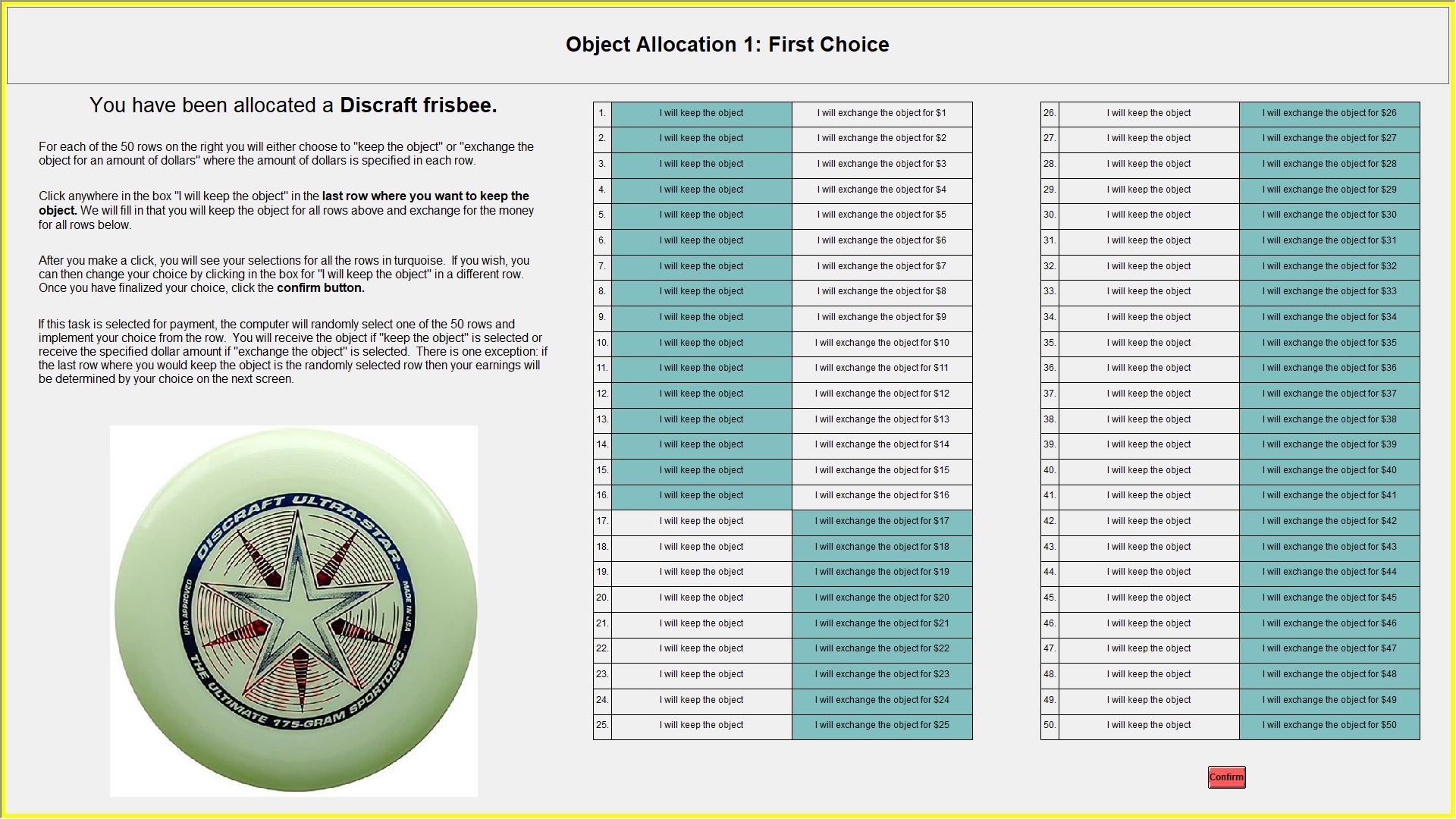}  
  \caption{First Screen}
  \label{fig:PhaseI1}
\end{subfigure}
\begin{subfigure}{\textwidth}
  \centering
  \includegraphics[width=\linewidth]{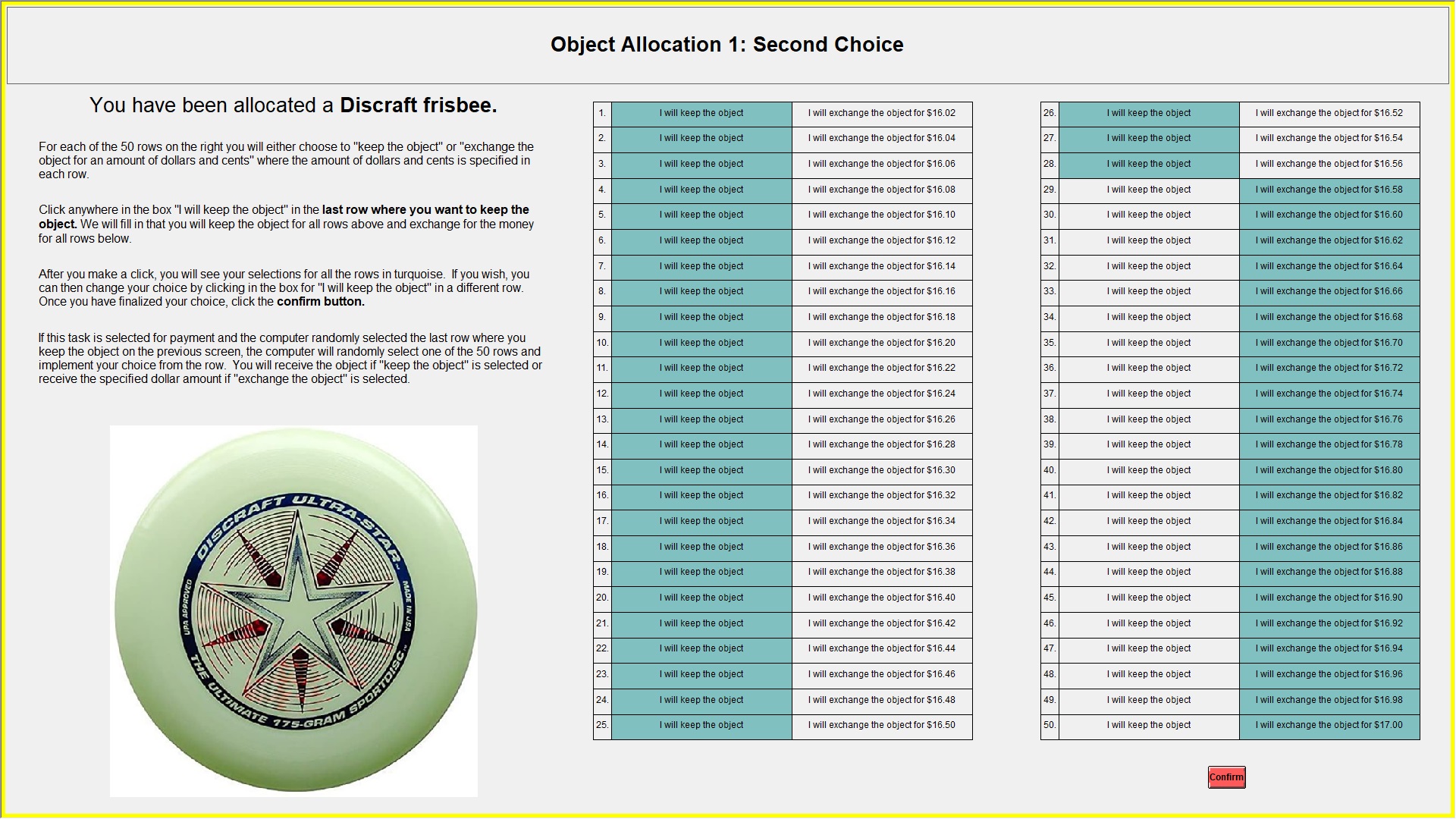}  
  \caption{Second Screen}
  \label{fig:PhaseI2}
\end{subfigure}
\caption{Value Elicitation Screen Shots}
\label{fig:PhaseISS}
\end{figure}

\newpage
\bibliographystyle{ecta}
\bibliography{references}

\end{document}

%% file: regressionsFormat.tex
{
\def\sym#1{\ifmmode^{#1}\else\(^{#1}\)\fi}
\begin{tabular}{l*{5}{c}}
\multicolumn{6}{c}{Table 1: Net Value Regressions}\\
\hline\hline
Treatment & \multicolumn{2}{c}{\emph{RSD}} & \multicolumn{2}{c}{\emph{Boston}} & All\\ \hline
            &\multicolumn{1}{c}{(1)}&\multicolumn{1}{c}{(2)}&\multicolumn{1}{c}{(3)}&\multicolumn{1}{c}{(4)}&\multicolumn{1}{c}{(5)}\\
\hline
rank   &     -0.8795*  &     -1.2358*  &      0.0979   &     -1.8063** &     -1.5526***\\
initial value&               &     -0.0570   &               &     -0.2768***&     -0.1759***\\
risk aversion&               &      0.0603   &               &      0.2676***&      0.1679***\\
loss aversion  &               &      0.0015   &               &     -0.1331   &     -0.0769   \\
CRT score    &               &      0.0222   &               &     -0.5717   &     -0.3484   \\
female      &               &     -0.1220   &               &     -0.6279   &     -0.5844   \\
Phase I order&               &      0.1077   &               &     -0.1333   &     -0.0328   \\
practice&               &     -0.1485   &               &     -0.2194   &     -0.1477   \\
truthful&               &     -1.6938   &               &     -0.7317   &     -1.1834   \\
\_cons      &      3.6033***&      4.5190   &      0.6334   &     10.4651** &      8.1221** \\
\hline
No. of Obs. &      100   &      100   &      100   &      100   &      200   \\
R-Squared   &        0.03   &        0.11   &        0.00   &        0.27   &        0.16   \\
\hline\hline
\end{tabular}
}

%% file: regressionstttFormat.tex
{
\def\sym#1{\ifmmode^{#1}\else\(^{#1}\)\fi}
\begin{tabular}{l*{6}{c}}
\multicolumn{7}{c}{Table 3: Truth-telling Top 2 Choices Regressions}\\
\hline\hline
Treatment & \multicolumn{2}{c}{\emph{RSD}} & \multicolumn{2}{c}{\emph{Boston}} & \multicolumn{2}{c}{All}\\ \hline
            &\multicolumn{1}{c}{(1)}&\multicolumn{1}{c}{(2)}&\multicolumn{1}{c}{(3)}&\multicolumn{1}{c}{(4)}&\multicolumn{1}{c}{(5)}&\multicolumn{1}{c}{(6)}\\
\hline
risk aversion&     -0.0160   &     -0.0117   &     -0.0255*  &     -0.0196   &     -0.0183*  &     -0.0147   \\
loss aversion  &      0.0051   &     -0.0053   &     -0.0060   &     -0.0013   &     -0.0004   &     -0.0029   \\
CRT score    &      0.2477*  &      0.2093   &      0.1546   &      0.0880   &      0.2011** &      0.1544   \\
female      &      0.5353   &      0.3808   &      0.5626** &      0.4994*  &      0.5189** &      0.4426** \\
practice&     -0.0135   &     -0.0166   &      0.0369   &      0.0255   &      0.0094   &      0.0034   \\
\_cons      &     -0.1425   &      0.5057   &      0.3021   &      0.4284   &      0.0348   &      0.4137   \\
\hline
Measure & Exact & \$2 diff. & Exact & \$2 diff. &Exact & \$2 diff. \\
No. of Obs. &      100   &      100   &      100   &      100   &      200   &      200   \\
\hline\hline
\end{tabular}
}

%% file: regressionsdumFormat.tex
{
\def\sym#1{\ifmmode^{#1}\else\(^{#1}\)\fi}
\begin{tabular}{l*{3}{c}}
\multicolumn{4}{c}{Table 5: Net Value Regressions with Dummies}\\
\hline\hline
Treatment & SD & Boston & All\\ \hline
            &\multicolumn{1}{c}{(1)}&\multicolumn{1}{c}{(2)}&\multicolumn{1}{c}{(3)}\\
\hline
ranked 2nd &      1.0655   &     -0.6298   &      1.2793   \\
ranked 3rd &      0.3480   &     -1.8964   &     -0.4721   \\
ranked 4th &      0.3046   &     -0.6471   &      0.4093   \\
ranked 5th &     -1.0711   &     -4.5459   &     -3.0656   \\
backpack &     10.3001***&      5.4340   &      8.1776***\\
notebook &      2.6014   &     -2.4099   &     -0.0261   \\
waterbottle &      7.5522***&      4.2488   &      6.3798***\\
pens &      1.7752   &     -4.0435*  &     -1.2110   \\
initial value&     -0.2130***&     -0.4093***&     -0.3186***\\
risk aversion&      0.0654   &      0.1927** &      0.1436***\\
\_cons      &     -1.6011   &      2.3024   &     -0.3960   \\
\hline
No. of Obs. &      100   &      100   &      200   \\
R-Squared   &        0.25   &        0.31   &        0.23   \\
\hline\hline
\end{tabular}
}

%% file: regressionsttFormat.tex
{
\def\sym#1{\ifmmode^{#1}\else\(^{#1}\)\fi}
\begin{tabular}{l*{6}{c}}
\multicolumn{7}{c}{Table 6: Truth-telling All Choices Regressions}\\
\hline\hline
Treatment & \multicolumn{2}{c}{RSD} & \multicolumn{2}{c}{Boston} & \multicolumn{2}{c}{All}\\ \hline

            &\multicolumn{1}{c}{(1)}&\multicolumn{1}{c}{(2)}&\multicolumn{1}{c}{(3)}&\multicolumn{1}{c}{(4)}&\multicolumn{1}{c}{(5)}&\multicolumn{1}{c}{(6)}\\
\hline
risk aversion&     -0.0224   &     -0.0142   &     -0.0211   &     -0.0270*  &     -0.0165   &     -0.0177*  \\
loss aversion  &      0.0000   &     -0.0057   &     -0.0224   &     -0.0042   &     -0.0110   &     -0.0046   \\
CRT score    &      0.1794   &      0.1956   &      0.2118   &      0.2475*  &      0.1863*  &      0.2250** \\
female      &      0.5718*  &      0.2012   &      0.4025   &      0.4182   &      0.4065*  &      0.3214   \\
practice&     -0.0701** &     -0.0507*  &      0.0481*  &      0.0282   &     -0.0024   &     -0.0105   \\
\_cons      &      0.1589   &      0.6542   &     -0.0824   &      0.0730   &     -0.0909   &      0.2440   \\
\hline
Measure & Exact & \$2 diff. & Exact & \$2 diff. &Exact & \$2 diff. \\
No. of Obs. &      100   &      100   &      100   &      100   &      200   &      200   \\
\hline\hline
\end{tabular}
}

%% file: regressionsttoFormat.tex
{
\def\sym#1{\ifmmode^{#1}\else\(^{#1}\)\fi}
\begin{tabular}{l*{6}{c}}
\multicolumn{7}{c}{Table 7: Truth-telling Top Choice Regressions}\\
\hline\hline
Treatment & \multicolumn{2}{c}{RSD} & \multicolumn{2}{c}{Boston} & \multicolumn{2}{c}{All}\\ \hline
            &\multicolumn{1}{c}{(1)}&\multicolumn{1}{c}{(2)}&\multicolumn{1}{c}{(3)}&\multicolumn{1}{c}{(4)}&\multicolumn{1}{c}{(5)}&\multicolumn{1}{c}{(6)}\\
\hline
risk aversion&     -0.0054   &     -0.0087   &     -0.0310*  &     -0.0225   &     -0.0150   &     -0.0138   \\
loss aversion  &      0.0076   &      0.0034   &     -0.0060   &     -0.0047   &      0.0004   &     -0.0008   \\
CRT score    &      0.2041   &      0.1716   &      0.0538   &      0.0787   &      0.1151   &      0.1197   \\
female      &      0.4395   &      0.3006   &     -0.0074   &      0.0962   &      0.1824   &      0.1757   \\
practice&     -0.0201   &     -0.0226   &     -0.0021   &     -0.0043   &     -0.0154   &     -0.0157   \\
\_cons      &      0.1928   &      0.7298   &      1.8629** &      1.6134** &      1.0138** &      1.1603** \\
\hline
Measure & Exact & \$2 diff. & Exact & \$2 diff. &Exact & \$2 diff. \\
No. of Obs. &      100   &      100   &      100  &      100   &      200   &      200   \\
\hline\hline
\end{tabular}
}